\documentclass[reprint,nobibnotes,superscriptaddress]{revtex4-1}
\usepackage{epsf}
\usepackage{epsfig}
\usepackage{amsmath}
\usepackage{amsfonts}
\usepackage{amssymb}
\usepackage{graphicx}
\usepackage{multirow}
\usepackage{graphicx}
\usepackage[table,xcdraw]{xcolor}
\usepackage{epsfig}
\usepackage{color}
\usepackage{bm}
\usepackage[yyyymmdd]{datetime}
\usepackage{dashrule}
\usepackage{ulem}
\usepackage{lineno}
\usepackage[margin=0.7in]{geometry}
\usepackage{xcolor}
\usepackage{ulem}

\begin{document}
	
\title{Detection of Acoustic Plasmons in Hole-Doped Lanthanum and Bismuth Cuprate Superconductors Using Resonant Inelastic X-Ray Scattering}

\author{Abhishek Nag}
\email[Email: ]{abhishek.nag@diamond.ac.uk}
\affiliation{Diamond Light Source, Harwell Campus, Didcot OX11 0DE, United Kingdom}
\author{M. Zhu}
\affiliation{H. H. Wills Physics Laboratory, University of Bristol, Bristol BS8 1TL, United Kingdom}
\author{Mat\'{\i}as Bejas}
\affiliation{Facultad de Ciencias Exactas, Ingenier\'{i}a y Agrimensura and Instituto de Física Rosario (UNR-CONICET), Av. Pellegrini 250, 2000, Rosario, Argentina}
\author{J. Li}
\affiliation{Diamond Light Source, Harwell Campus, Didcot OX11 0DE, United Kingdom}
\affiliation{Beijing National Laboratory for Condensed Matter Physics, Institute of Physics, Chinese Academy of Sciences, Beijing 100190, China}
\author{H. C. Robarts}
\affiliation{Diamond Light Source, Harwell Campus, Didcot OX11 0DE, United Kingdom}
\affiliation{H. H. Wills Physics Laboratory, University of Bristol, Bristol BS8 1TL, United Kingdom}
\author{Hiroyuki Yamase}
\affiliation{International Center of Materials Nanoarchitectonics, National Institute for Materials Science, Tsukuba 305-0047, Japan}
\affiliation{Department of Condensed Matter Physics, Graduate School of Science, Hokkaido University, Sapporo 060-0810, Japan}
\author{A. N. Petsch}
\affiliation{H. H. Wills Physics Laboratory, University of Bristol, Bristol BS8 1TL, United Kingdom}
\author{D. Song}
\affiliation{National Institute of Advanced Industrial Science and Technology (AIST), Tsukuba, Ibaraki 305-8560, Japan}
\author{H. Eisaki}
\affiliation{National Institute of Advanced Industrial Science and Technology (AIST), Tsukuba, Ibaraki 305-8560, Japan}
\author{A. C. Walters}
\affiliation{Diamond Light Source, Harwell Campus, Didcot OX11 0DE, United Kingdom}
\author{M. Garc\'{\i}a-Fern\'{a}ndez}
\affiliation{Diamond Light Source, Harwell Campus, Didcot OX11 0DE, United Kingdom}
\author{Andr\'{e}s Greco}
\affiliation{Facultad de Ciencias Exactas, Ingenier\'{i}a y Agrimensura and Instituto de Física Rosario (UNR-CONICET), Av. Pellegrini 250, 2000, Rosario, Argentina}
\author{S. M. Hayden}
\email[Email: ]{s.hayden@bristol.ac.uk}
\affiliation{H. H. Wills Physics Laboratory, University of Bristol, Bristol BS8 1TL, United Kingdom}
\author{Ke-Jin Zhou}
\email[Email: ]{kejin.zhou@diamond.ac.uk}
\affiliation{Diamond Light Source, Harwell Campus, Didcot OX11 0DE, United Kingdom}


\date{\today}

\begin{abstract}
High $T_c$ superconductors show a rich variety of phases associated with their charge degrees of freedom. Valence charges can give rise to charge ordering or acoustic plasmons in these layered cuprate superconductors. While charge ordering has been observed for both hole- and electron-doped cuprates, acoustic plasmons have only been found in electron-doped materials. Here, we use resonant inelastic X-ray scattering (RIXS) to observe the presence of acoustic plasmons in two families of hole-doped cuprate superconductors (La$_{1.84}$Sr$_{0.16}$CuO$_4$ and Bi$_2$Sr$_{1.6}$La$_{0.4}$CuO$_{6+\delta}$), crucially completing the  picture. Interestingly, in contrast to the quasi-static charge ordering which manifests at both Cu and O sites, the observed acoustic plasmons are predominantly associated with the O sites, revealing a unique dichotomy in the behaviour of valence charges in hole-doped cuprates. 
\end{abstract}	
	
\maketitle

\begin{figure*}[tbph!]
	\begin{center}
		\includegraphics{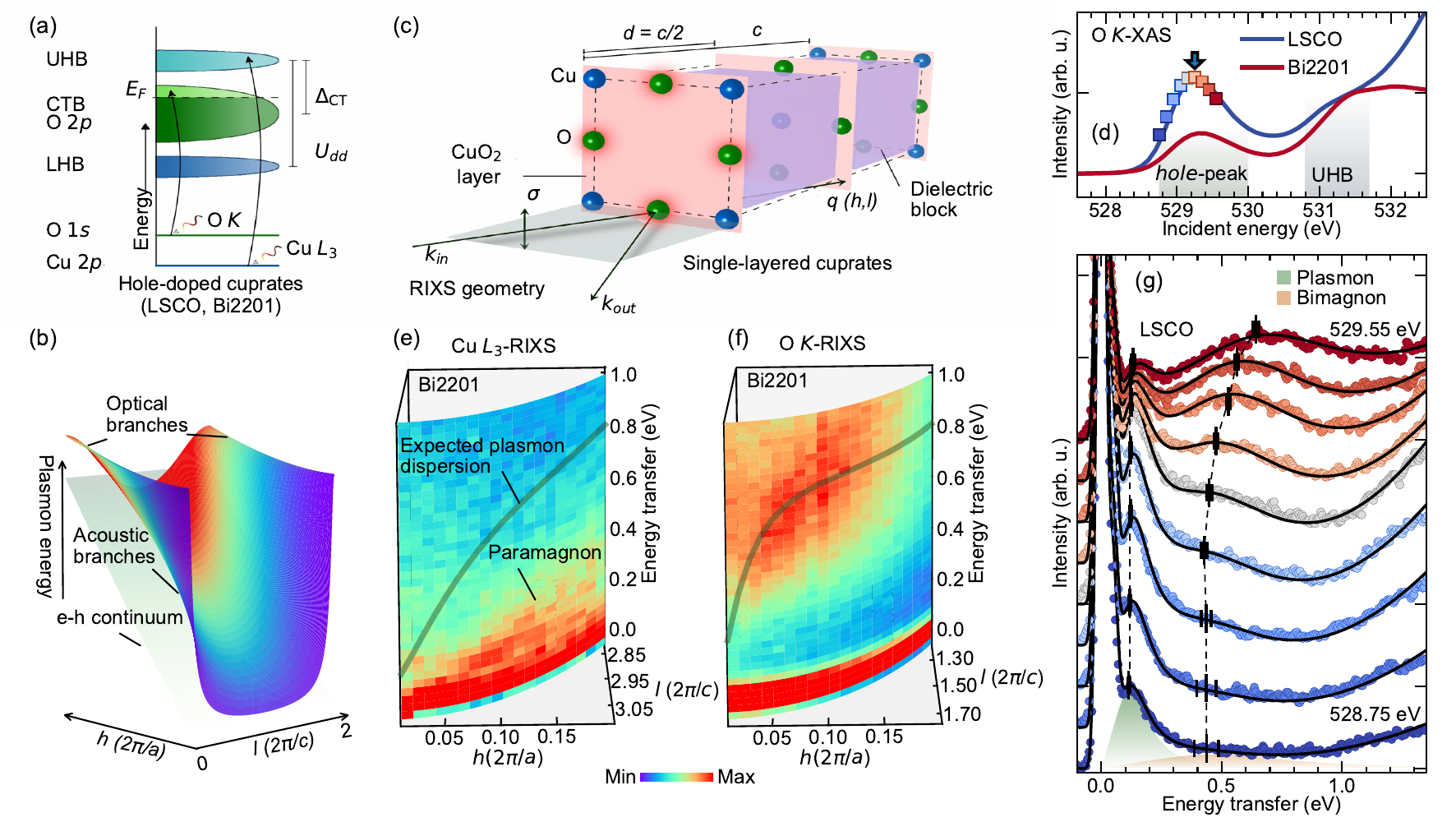}	
		\caption{(a) Schematic electronic stucture of hole-doped cuprates in the Zaanen-Sawatzky-Allen scheme~\cite{zaanen1985prl,emery1987prl,varma1987ssc,armitage2010rmp}. (b) Acoustic plasmon branches dispersing towards zero-energy in the $h$-direction while maintaining a periodicity of $l=2$ in the $l$-direction for the layered structure shown in panel (c). (c) RIXS scattering geometry and a representative unit cell of single-layered cuprates. (d) Peak feature attributed to the doped-holes observed through the O $K$-XAS process~\cite{chen1991prl}. The arrow shows the incident energy chosen to probe the plasmon dispersions in this work.  RIXS intensity maps for Bi2201 at a fixed scattering angle of 114$^{\circ}$ at (e) Cu $L_3$-  and (f) O $K$-edges. (g) Vertically stacked RIXS spectra for LSCO at  ($h=0.03$, $l=1.00$). Colours of the spectra correspond to incident energies shown in (d). Solid black lines are fits to the RIXS spectra. Vertical black bars are least-square-fit peak positions.
		} \label{fig1}
	\end{center}
\end{figure*}

\noindent
The electronic structure of high temperature superconducting layered-cuprates~\cite{keimer2015nat} may be understood in terms of a hybridisation between the Cu $3d_{x^2-y^2}$ and O $2p_{\sigma}$ orbitals, and a strong on-site Coulomb repulsion between electrons on the Cu sites~\cite{zaanen1985prl,emery1987prl,varma1987ssc,armitage2010rmp}. When holes are introduced (see Fig.~\ref{fig1}(a)), they reside preferentially in the so-called ``charge-transfer band'' (CTB) which is composed primarily of O orbitals~\cite{chen1991prl}. In contrast, doped electrons enter the upper Hubbard band (UHB) and primarily reside on the Cu sites~\cite{armitage2010rmp}. Despite this asymmetry in the electronic structure, charge order, a complex phase of periodically modulated charge-carrier density, has been observed ubiquitously on both the electron- and hole-doped sides of the phase diagram~\cite{comin2016arcmp}.

Surprisingly, a more widely observed mode of collective charge density oscillation, the acoustic plasmon~\cite{diaconescu2007nat}, has been rather elusive for the cuprates. In contrast to three-dimensional (3D) metals, where long-range Coulomb interactions give rise to isotropic long-wavelength optical-like gapped plasmons, out-of-phase oscillations of charges in neighbouring planes of two-dimensional (2D) layered electron gases, form acoustic plasmons, whose energy tends to zero for small in-plane wavevectors (see Fig.~\ref{fig1}(b))~\cite{fetter1973ap,fetter1974ap}. Due to confinement of the doped-charges to CuO$_2$ planes and poor screening of out-of-plane Coulomb interactions by intervening dielectric blocks (see Fig.~\ref{fig1}(c)), acoustic plasmons are also expected in the layered cuprates~\cite{kresin1988prb,greco2019cp,markiewicz2008prb}.

A remarkable discovery has been the recent observation of acoustic plasmons in electron-doped La$_{2-x}$Ce$_x$CuO$_4$ (LCCO) and Nd$_{2-x}$Ce$_x$CuO$_4$~\cite{hepting2018nat,lin2020npj} using Cu $L_{3}$-RIXS. The excitations were found to have the strong out-of-plane dispersion expected for plasmons in layered systems. The situation in hole-doped cuprates, however, has remained rather controversial.  While Cu $L_3$-RIXS did not detect plasmons in several families~\cite{lee2014np,dellea2017prb,miao2017pnas}, O $K$-RIXS did detect excitations in La$_{2-x}$(Br,Sr)$_x$CuO$_4$ that were interpreted as incoherent intra-band transitions~\cite{ishii2017prb}. In zero-momentum optical investigation of Bi$_2$Sr$_2$CaCu$_2$O$_8$, apart from observation of optical plasmons at 1.1 eV, a low-energy non-Drude behaviour was contemplated to be due to a band of acoustic plasmons~\cite{bozovic1990prb}. However, electron energy-loss spectroscopy, a traditional probe for studying plasmons, observed only a high energy ($\sim$ 1~eV) overdamped optical plasmon for small in-plane wavevectors in  Bi$_{2.1}$Sr$_{1.9}$CaCu$_{2}$O$_{8+\delta}$~\cite{mitrano2018pnas}. Acoustic plasmons in layered systems originate from the presence of conduction electrons and the long-range nature of the Coulomb interaction. Their absence in hole-doped cuprates would conflict with our general understanding of the collective behaviour of the doped-charges.

In this letter we report that acoustic plasmons are indeed present in hole-doped cuprates from an extensive O $K$-RIXS study of  La$_{1.84}$Sr$_{0.16}$CuO$_4$ (LSCO) and Bi$_2$Sr$_{1.6}$La$_{0.4}$CuO$_{6+\delta}$ (Bi2201) over a wide range of in- and out-of-plane momenta. The discovery of acoustic plasmons in the hole-doped systems remarkably illustrates the universal existence of low-energy collective excitations besides phonons and spin-fluctuations across the cuprate phase diagram. Surprisingly, the observed acoustic plasmons are predominantly associated with the O sites in these hole-doped cuprates. Our results will therefore stimulate more studies of doped-hole charge dynamics, taking into account the three band model in the cuprates~\cite{varma1987ssc,emery1987prl}.

Spectroscopically, Cu $L_3$- and O $K$-RIXS directly probe the charge and magnetic excitations associated with the Cu 3$d$ and O 2$p$ orbitals, respectively, at the corresponding absorption peaks (Fig.~\ref{fig1}(a)). In order to compare the excitations associated with the two orbitals, high-resolution RIXS spectra were collected at Cu $L_3$- ($\bigtriangleup E\simeq$ 0.045 eV) and O $K$- ($\bigtriangleup E\simeq$ 0.043 eV) edges, at I21-RIXS beam line, Diamond Light Source, United Kingdom~\cite{i21web,SM}, in the scattering geometry shown in Fig.~\ref{fig1}(c). All data presented here were obtained with incident $\sigma$ polarisation (perpendicular to the scattering plane) to enhance the charge excitations~\cite{hepting2018nat}. Single crystals of LSCO and Bi2201 were cooled to their respective $T_c$s of 38~K and 34.5~K,consistent with optimal hole-doping of $p=0.16$, and X-ray absorptions (XAS) were collected in total electron yield mode (see Fig.~\ref{fig1}(d))~\cite{SM}. A survey was first made near the in-plane zone-centre with a fixed scattering angle on Bi2201 (Fig.~\ref{fig1}(e, f)). The low-energy inelastic spectra at Cu $L_3$ resonance are dominated by paramagnons without any noticeable signs of plasmons, similar to reports on other hole-doped systems~\cite{lee2014np,dellea2017prb,miao2017pnas}. At the O $K$-edge \textit{hole}-peak~\cite{chen1991prl} however, a mode is found below 1 eV, dispersing towards the zero-energy. 

\begin{figure*}[hbt!]
	\begin{center}
		\includegraphics{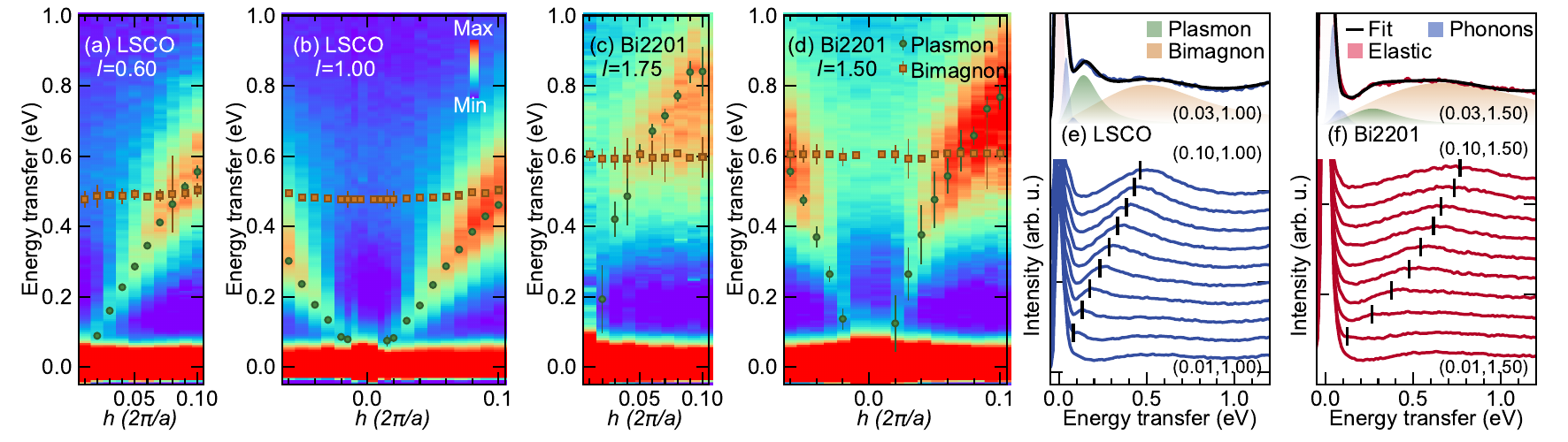}
		\caption{RIXS intensity maps of LSCO for momentum transfer along the $h$-direction at (a) $l=0.60$ and (b) $l=1.00$ and of Bi2201 at (c) $l=1.75$ and (d) $l=1.50$. (e,f) Top, representative RIXS spectra at mentioned ($h$, $l$)-values for LSCO and Bi2201. Shaded areas represent the different peak profiles~\cite{SM}. (e,f) Bottom, vertically stacked RIXS spectra from $h=0.01$ to 0.10 for LSCO and Bi2201 at $l=1.00$ and $l=1.50$, respectively.
		} \label{fig2}
	\end{center}
\end{figure*}
	
We next collected RIXS spectra by varying the incident energy ($E_i$) across the hole-peak in the O $K$-XAS of LSCO at ($h=0.03$, $l=1.00$) (Fig.~\ref{fig1}(d))~\cite{chen1991prl}, as shown in Fig.~\ref{fig1}(g). We denote momentum transfers along $h$-, $k$- and $l$-directions in reciprocal lattice units, where $\mathbf{Q}=ha^*+kb^*+lc^*$ ($a^*=2\pi/a$, $b^*=2\pi/b$, $c^*=2\pi/c$), $k=0$ if not stated explicitly. We find a broad feature at $\sim0.5$~eV shifting towards higher energies with increasing $E_i$. With doping, the probability of scattering from doped-charges in the intermediate state of RIXS increases~\cite{bisogni2012prb}. Moreover, energy-shift of the magnetic excitations associated with incoherent charge excitations is enhanced in $\sigma$-polarised RIXS~\cite{minola2017prl}. Thus, this feature can be ascribed to bimagnon excitations with an itinerant character~\cite{bisogni2012prb}.We find an additional sharp mode, at $\sim0.13$~eV, whose energy remains constant with $E_i$. This is a signature of its coherent nature~\cite{lin2020npj} and is in contrast with previous O $K$-RIXS results~\cite{ishii2017prb}. This feature cannot be due to two-particle electron-hole like excitations, which are incoherent in nature. Neither can it be due to single-magnons or paramagnons since $\Delta S = 1$ spin-flip processes are forbidden at the O $K$-edge~\cite{bisogni2012prb}. To ascertain its origin, we explored further its dispersion in energy-momentum space.

The broad feature seen in Fig.~\ref{fig1}(g) is almost non-dispersive in the $h$-direction further confirming its assignment as bimagnons~\cite{bisogni2012prb}. This can be seen from $(h,E)$-maps collected at constant $l$-values for LSCO and Bi2201 (Fig.~\ref{fig2}). In contrast, the sharp mode disperses towards zero energy near the in-plane zone-centre in both systems. The RIXS spectra were fitted with a sum of Gaussians for the elastic line, phonons and damped harmonic oscillator functions for the sharp mode and the broad feature (Fig.~\ref{fig2}(e-f))~\cite{SM}. The dispersion and reduction of amplitude and width towards the in-plane zone-centre of this mode, is reminiscent of the acoustic plasmon behaviour observed in electron-doped LCCO~\cite{hepting2018nat}. The spectral weight of this mode moves to lower energy as the doping is reduced (see Supplementary Materials Fig. S15~\cite{SM}). This is expected and consistent with the behaviour of plasmons in LCCO~\cite{hepting2018nat,lin2020npj}. However, the mode is strongly damped in comparison to LCCO (see Supplementary Materials Fig.~S12~\cite{SM}), reflecting the stronger correlations (e.g. pseudogap) near optimal doping in hole-doped cuprates~\cite{armitage2010rmp,lin2020npj}.

\begin{figure*}[hbt!]
	\begin{center}
		\includegraphics{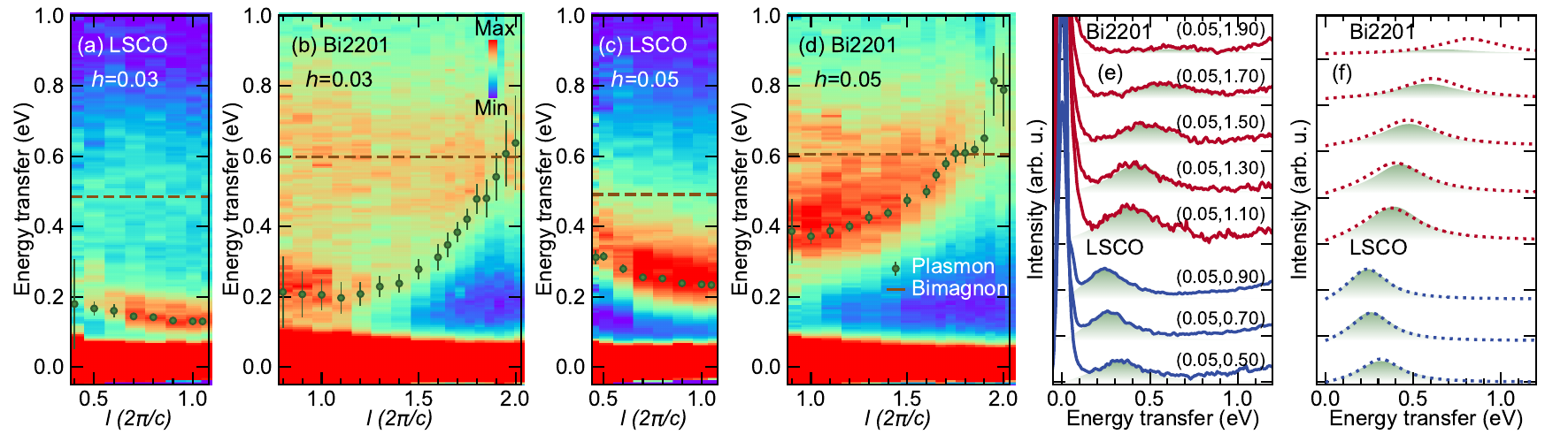}
		\caption{RIXS intensity maps of LSCO and Bi2201 for momentum transfer along the $l$-direction at (a, b) $h=0.03$ and (c, d) $h=0.05$. (e) Representative RIXS spectra at mentioned ($h$, $l$)-values for LSCO (blue lines) and Bi2201 (red lines). Fitted peak profiles of bimagnons have been subtracted from the RIXS spectra to show the evolution of the plasmon peak (f) $\mathbf{\chi}_c''$($\mathbf{Q}$, $\omega$) calculated from $t$-$J$-$V$ model for corresponding ($h$, $l$)-values in (e) for LSCO (Bi2201) are shown by the dotted blue (red) lines. Shaded green areas are the plasmon peak profiles obtained from RIXS data fitting. Different broadening factors  ($\Gamma=0.29t$ for Bi2201 and $0.2t$ for LSCO) were chosen to replicate the lineshapes of the two materials. 
		} \label{fig3}
	\end{center}
\end{figure*}

The most stringent test for identifying the mode as plasmons in these systems is their $l$-dispersion. In the out-of-plane direction, plasmons in layered electron systems have a periodicity of $2\pi/d$ (where $d$ is the interlayer spacing), which corresponds to $l=2$  in these systems, with a minimum in energy at $l=1,3,5...$ (Fig.~\ref{fig1}(c)). LSCO and Bi2201 have interlayer spacings  differing by a factor of $\sim$2, allowing us to probe separate portions of this period. The sharp mode observed in Fig.~\ref{fig2}, is found to disperse to a minimum energy value at $l = 1$  for both systems.  This can be seen in the ($l, E)$-maps collected at fixed $h$-values shown in Fig.~\ref{fig3}(a-d). This behaviour fundamentally proves the presence of acoustic plasmons in hole-doped cuprates. We can exclude the previous interpretation of these excitations as incoherent intra-band charge or electron-hole excitations which are 2D~\cite{ishii2017prb}, without significant $l$-dependence~\cite{greco2019cp}. We note that a limited out-of-plane dispersion study has also been done recently on underdoped LSCO~\cite{singh2020arx}.

The cuprates are strongly correlated electron systems~\cite{keimer2015nat}. As such, it is interesting to compare our experimental results with the recently developed calculations of plasmons in the framework of a $t$-$J$-$V$ model~\cite{greco2019cp}, although generic plasmon behaviour can also be described by random-phase-approximation calculations~\cite{kresin1988prb,markiewicz2008prb}. For discussing the nature and origin of the 3D charge excitations in LSCO and Bi2201 we employed the minimal layered $t$-$J$-$V$ model~\cite{greco2019cp,zhang1988prb,SM}:

\begin{eqnarray}
H = &&-\sum_{i, j,\sigma} t_{i j}\tilde{c}^\dag_{i\sigma}\tilde{c}_{j\sigma}
+ \sum_{\langle i,j \rangle} J_{ij} \left( \vec{S}_i \cdot \vec{S}_j - \frac{1}{4} n_i n_j \right)\nonumber \\
&&+\frac{1}{2} \sum_{i,j}  V_{ij} n_i n_j. 
\label{tJV_main} 
\end{eqnarray}

\noindent Here, $t_{i j}$ represents the hopping parameter and $J_{i j}$ the exchange parameter.  The 3D form of long-range Coulomb interaction $V_{i j}$ used in Eq.~\ref{tJV_main} in momentum space is~\cite{becca1996prb}:
\begin{equation}
V(\mathbf{Q})=\frac{V_c}{A(q_x,q_y) - \cos q_z},
\label{LRC_main}
\end{equation}

\noindent where $V_c= e^2 d(2 \epsilon_{\perp} a^2)^{-1}$ and $A(q_x,q_y)= \alpha (2 - \cos q_x - \cos q_y)+1$ 
with $\alpha=\frac{\epsilon_\parallel/\epsilon_\perp}{(a/d)^2}$, $e$ the elementary charge and high frequency in-($\epsilon_{\parallel}$) and out-of-plane ($\epsilon_{\perp}$) dielectric constants.

\begin{figure}[tbph!]
	\begin{center}
		\includegraphics{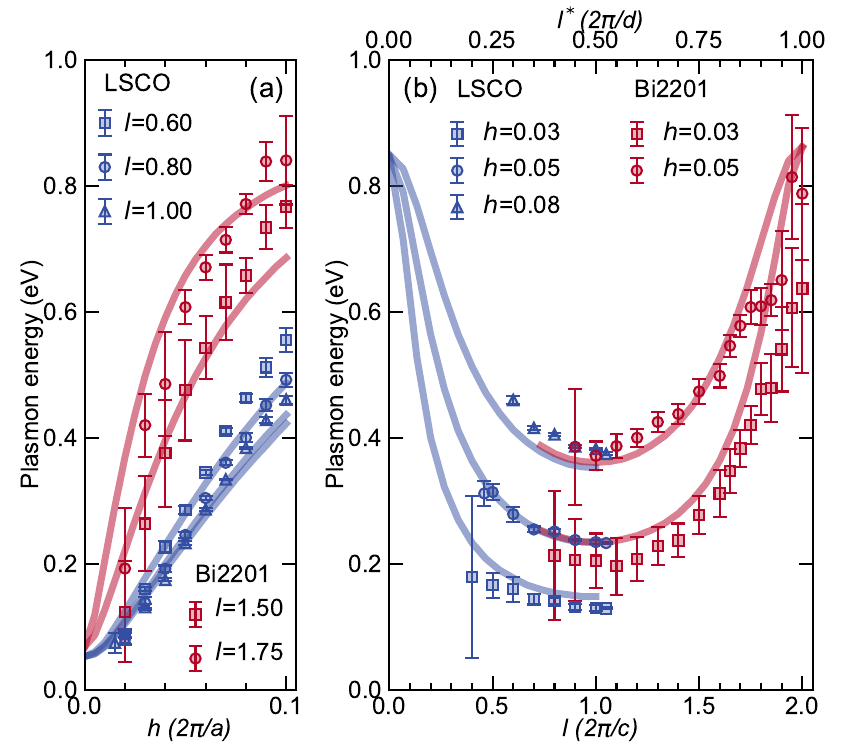}
		\caption{(a) Plasmon energies in LSCO and Bi2201 for momentum transfer along the $h$-direction for different $l$-values and (b) for momentum transfer along the $l$-direction for different $h$-values, summarised from least-square-fits of RIXS spectra.Continuous lines are plasmon dispersions obtained from $t$-$J$-$V$ model independently optimised for LSCO (blue) and Bi2201 (red).
		} \label{fig4}
	\end{center}
\end{figure}

The imaginary part of  charge susceptibility $\mathbf{\chi}_c''$($\mathbf{Q}$, $\omega$), obtained from the model resembles well the spectral shape of the plasmons for both systems, as shown in Fig.~\ref{fig3}(f) at different ($h$, $l$)-values. This demonstrates that the charge excitations in RIXS, although influenced by resonance and polarisation effects, can fundamentally be related to the charge-density response function. At $l$-values close to 2, the much larger suppression of plasmons compared to theory could be due to their decay through the incoherent charge channels associated with bimagnons or through Umklapp scattering~\cite{hepting2018nat}. Severe suppression of the charge excitations in this region forbids us from detecting the optical plasmon branch in Bi2201 (see Supplementary Materials Fig.~S6)~\cite{SM}. In Fig.~\ref{fig4}(a, b) we show that both the $h$- and $l$-direction plasmon dispersions extracted by fitting the RIXS spectra for LSCO and Bi2201 are also represented well by the $t$-$J$-$V$ model optimised independently for each material. The higher acoustic plasmon velocities in Bi2201 than LSCO seen in Fig.~\ref{fig4}(a, b), arise mainly  due to the larger interlayer spacing, considering the two systems have similar carrier densities and Fermi velocities (see Supplementary Materials Section V)~\cite{fetter1973ap,hepting2018nat,SM}. Nevertheless, the remarkable similarity of the plasmon dispersions in the two different families suggests their ubiquitous existence in hole-doped cuprates. 

In order to shed light on the possible existence of plasmons at Cu sites, we use the model described above to calculate the expected plasmon energies for both systems along the ($h$, $l$)-paths corresponding to the data for fixed scattering geometry. Neither in Bi2201 (Fig.~\ref{fig1}(e)) nor in LSCO (Supplementary Materials Fig.~S2)~\cite{SM}), there is any spectral weight evident in Cu $L_3$-RIXS spectra that can be assigned to plasmons~\cite{lee2014np,dellea2017prb}. This contrasts with the strong plasmon signals observed at the O sites. The amplitude of plasmon is found to be highest close to $l=1$ for the O $K$-RIXS (Fig.~\ref{fig3} and Supplementary Materials Fig.~S12(c)~\cite{SM}). In the Cu $L_3$-RIXS experiments close to the in-plane zone-centre, we probe near $l=3$ for Bi2201 and near $l=1.8$ for LSCO. Plasmons were however, clearly observed in Cu $L_3$-RIXS of electron-doped LCCO at similar $l$-values as in LSCO~\cite{hepting2018nat}. The non-observance of plasmons at Cu sites in the present study of LSCO and Bi2201 could be therefore due to a combined effect of an $l$-dependence of plasmon spectral weight and a strong O 2$p$ character of the doped-charges~\cite{chen1991prl,zhang1988prb,sakurai2011sci}. Further studies are required to clarify such site-dependent behaviour, since in the Zhang-Rice singlet state scenario for hole-doped cuprates, a strong coupling is expected between the doped holes in the O $2p_{\sigma}$ orbitals and the intrinsic holes of the Cu $3d_{x^2-y^2}$ orbitals~\cite{zhang1988prb}.

It is interesting to discuss our findings in the context of the charge order type of density modulation observed in hole-doped cuprates. When charge order is present, it has been observed using both Cu $L_3$- and O $K$-edge resonant X-ray scattering ~\cite{abbamonte2005np,achkar2013prl,li2020pnas}, both for short ($\sim15$ \AA)~\cite{li2020pnas} and long ($\sim200$ \AA)~\cite{abbamonte2005np} correlation lengths. In common with other charge-density waves, the order has a valence charge modulation and associated atomic displacements \cite{johannes2008prb}, making it possible to be observed by non-resonant X-ray scattering techniques~\cite{forgan2015natcomm}. Thus, it is likely that, in these systems, the charge ordering signal observed at the Cu $L_3$ absorption peak is primarily due to atomic displacements caused by electron-phonon coupling, while the dominant signal at the O $K$ hole-peak reflects directly the valence charge modulation \cite{achkar2013prl,fink2009prbr}. Due to the much higher frequencies of the dynamic plasmons, it may be that they couple weakly to the phonons, further reducing the possibility to observe any signature directly from the Cu 3$d$ orbitals.


The general existence of acoustic plasmons besides phonons and spin-fluctuations in layered cuprates will lead to more investigations of charge dynamics in connection with the pseudogap phase, non-Fermi liquid behaviour and perhaps the superconductivity in cuprates~\cite{keimer2015nat}. Our results also suggest that the charge dynamics in hole-doped cuprates are mostly associated with the O sites, highlighting the importance of the three band model in the cuprates~\cite{varma1987ssc,emery1987prl}. Going beyond, it would be interesting to utilise the site-sensitivity of the RIXS technique to characterise plasmon behaviour in other layered superconductors, like iron-pnictides~\cite{paglione2010nat}, having strong out-of-plane band dispersions, or the newly-found nickelates in which 2D Ni-3$d$ states strongly hybridise with 3D rare-earth 5$d$ states~\cite{hepting2020nm}.

\begin{acknowledgements}
We thank W.-S. Lee, V. Kresin, J. Lorenzana and S. Johnston for insightful discussions. All data were taken at the I21 RIXS beamline of Diamond Light Source (United Kingdom) using the RIXS spectrometer designed, built and owned by Diamond Light Source. We acknowledge Diamond Light Source for providing the beamtime on Beamline I21 under proposals SP20709 and MM24587. Work at Bristol was supported by EPSRC Grant EP/R011141/1 and EP/L015544/1. H. Y. was supported by JSPS KAKENHI Grant No. JP18K18744 and JP20H01856.  We acknowledge T. Rice for the technical support throughout the beamtimes. We also thank G. B. G. Stenning and D. W. Nye for help on the Laue instrument in the Materials Characterisation Laboratory at the ISIS Neutron and Muon Source. 
\end{acknowledgements}

\pagebreak
\onecolumngrid

\begin{center}
	\textbf{\large Supplementary Materials}\\[.2cm]
\end{center}

\setcounter{equation}{0}
\setcounter{figure}{0}
\setcounter{table}{0}
\setcounter{page}{1}

\newcommand{\hbAppendixPrefix}{S}
\renewcommand{\thefigure}{\hbAppendixPrefix\arabic{figure}}	
\renewcommand{\theequation}{\hbAppendixPrefix\arabic{equation}}	

\section{Sample growth, characterisation and preparation.}
\begin{figure}[ht!]
	\begin{center}
		\includegraphics{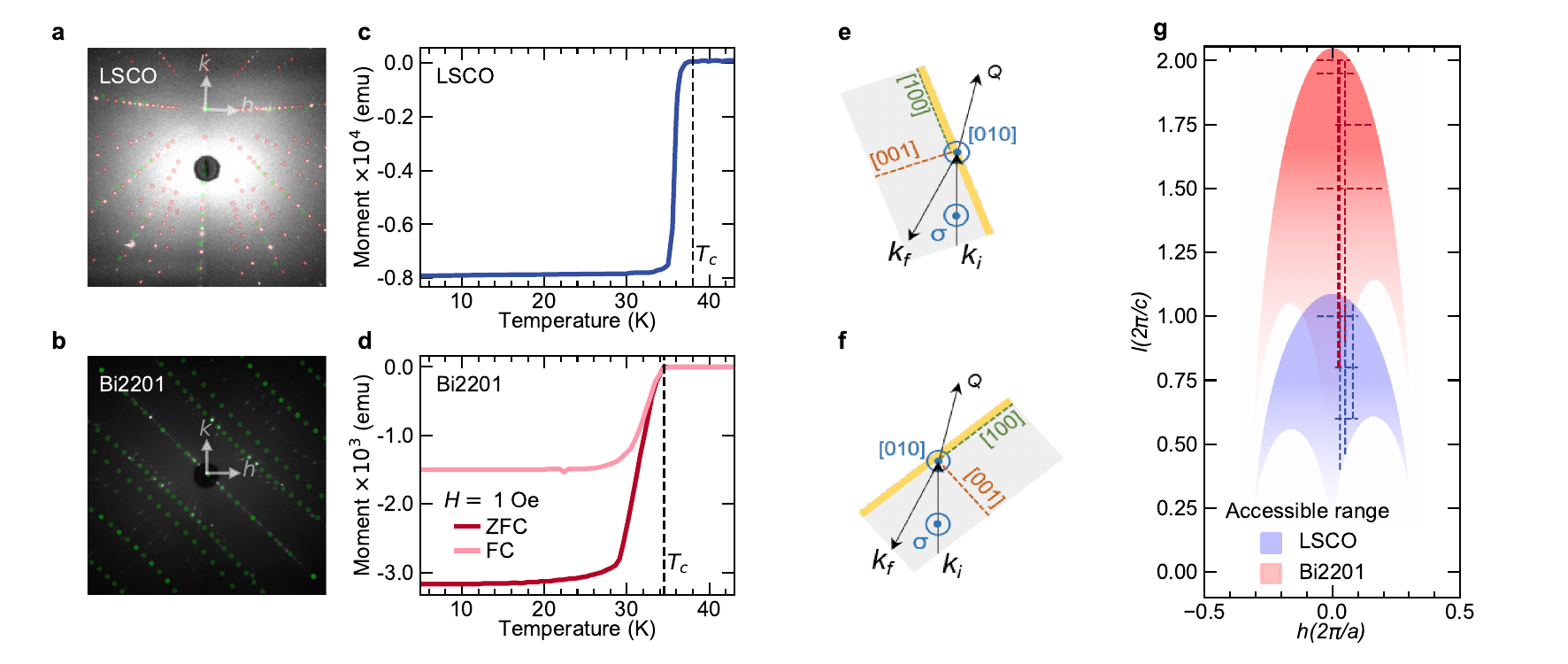}
		\caption{\textbf{Sample properties.}
			textbf{a, b} Laue diffraction patterns of LSCO and Bi2201 samples. \textbf{c, d} Zero-Field-Cooled (ZFC) and Field-Cooled (FC) magnetisation curves for LSCO and Bi2201. The onset of the superconducting transition $T_c$ is shown by the vertical dotted lines. \textbf{e} Grazing-incident and \textbf{f} grazing-exit scattering geometries. \textbf{g} Shows the accessible ($h$, $l$)-values at O $K$-for LSCO and Bi2201 at I21. Negative and positive values of $h$ represent the grazing-in and grazing-exit configurations respectively. Vertical and horizontal lines show the ($h$-, $l$)-trajectories along which RIXS spectra were collected in this work.
		} \label{Sample_info}
	\end{center}
\end{figure}

\noindent
High-quality single crystals of La$_{1.84}$Sr$_{0.16}$CuO$_4$ (LSCO) and Bi$_2$Sr$_{1.6}$La$_{0.4}$CuO$_{6+\delta}$ (Bi2201) were grown by floating-zone method. Bi2201 sample was annealed at 650 $^{\circ}$C in O$_2$ atmosphere for two days to improve sample homogeneity. Fig.~\ref{Sample_info}(a) and (b) show the Laue diffraction patterns obtained from LSCO and Bi2201 respectively, for preliminary orientation. The orientations were further refined using in-situ diffraction, charge-order and superstructure (present in Bi2201 due to structural distortions in Bi-O layers) peaks prior to the collection of RIXS spectra. Lattice constants used for LSCO (Bi2201) are $a=b=3.77$ (3.86) \AA~and $c=13.1$ (24.69) \AA. Superconducting transition temperatures $T_c$s extracted from magnetisation measurements of LSCO (38~K) and Bi2201 (34.5~K) are shown in Fig.~\ref{Sample_info}(c) and (d) and are consistent with optimal hole-doping of $p$ =0.16. While LSCO was cleaved in vacuum, Bi2201 was cleaved in air and immediately transferred to vacuum. The pressure inside the sample vessel was maintained at $\sim5\times10^{-10}$ mbar. Samples were mounted such that the $a$-axis and $c$-axis lay in the horizontal scattering plane while the $b$-axis was perpendicular to the scattering plane (see Fig.~1(c) of Main paper). Negative and positive values of $h$ in the RIXS maps presented in this work represent the grazing-incident and grazing-exit geometries respectively (Fig.~\ref{Sample_info}(e, f)). Fig.~\ref{Sample_info}(g) shows the accessible ($h$, $l$)-values at O $K$-for LSCO and Bi2201 at I21. For each material, reduction in $l$-value forces transition from a backward to forward scattering experimental geometry which enhances the elastic line, thereby limiting the lowest usable $l$-to study plasmons. The zero-energy transfer position and resolution of the RIXS spectra were determined from subsequent measurements of elastic peaks from an adjacent carbon tape.

\section{Additional RIXS data}
\begin{figure}[htpb!]
	\begin{center}
		\includegraphics[width=\textwidth]{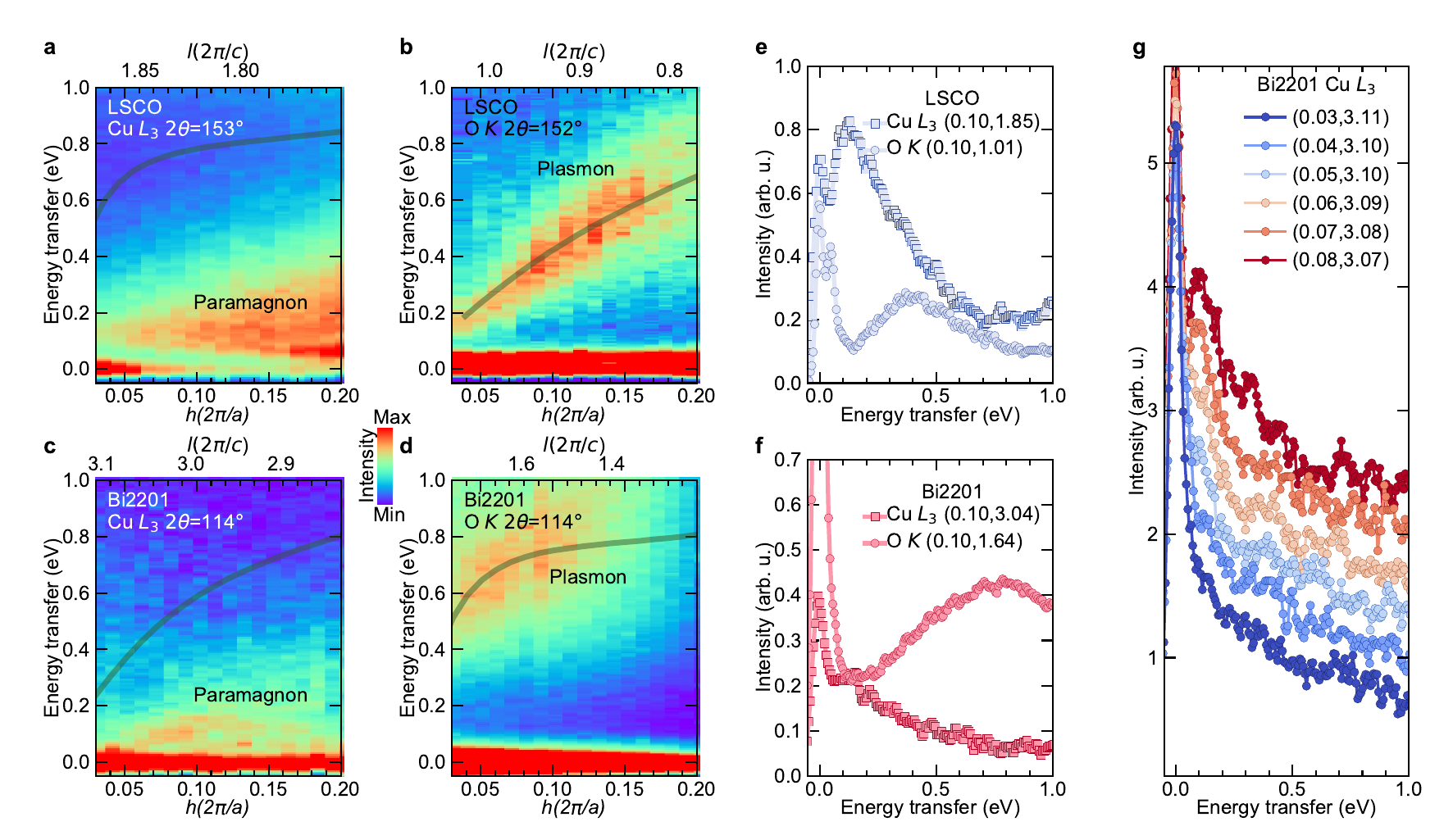}
		\caption{\textbf{Comparison of Cu $L_3$-and O $K$-RIXS for probing plasmons in LSCO and Bi2201.}
			\textbf{a, b} RIXS intensity maps of LSCO for momentum transfer along the $h$-direction and \textbf{c, d} of Bi2201 at Cu $L_3$ absorption and  O $K$ hole-peaks. The spectra are collected with a fixed scattering angle (2$\theta$) for each map and hence the ($h, l$)-values vary simultaneously. Green solid lines show the expected plasmon dispersions from $t$-$J$-$V$ model. \textbf{e, f} Representative RIXS spectra for LSCO and Bi2201 at given ($h$, $l$)-values for Cu $L_3$- (square symbols) and  O $K$- (circle symbols) edges. \textbf{g} RIXS spectra for Bi2201 at given ($h$, $l$)-values for Cu $L_3$-edge.   
		} \label{Edge_comparison_v2}
	\end{center}
\end{figure}

\noindent
RIXS spectra were collected at both Cu $L_3$-and O $K$-RIXS for comparison of plasmons in LSCO and Bi2201 with a fixed scattering angle. Data collected from LSCO at the Cu $L_3$ absorption peak is shown in Fig.~\ref{Edge_comparison_v2}(a) and at the hole-peak of O $K$-edge in Fig.~\ref{Edge_comparison_v2}(b).  Data presented in Fig.~1 (e, f) of main paper on Bi2201 are also presented for a closer inspection in Fig.~\ref{Edge_comparison_v2}(c, d, f and g). 

\begin{figure*}[htbp!]
	\begin{center}
		\includegraphics{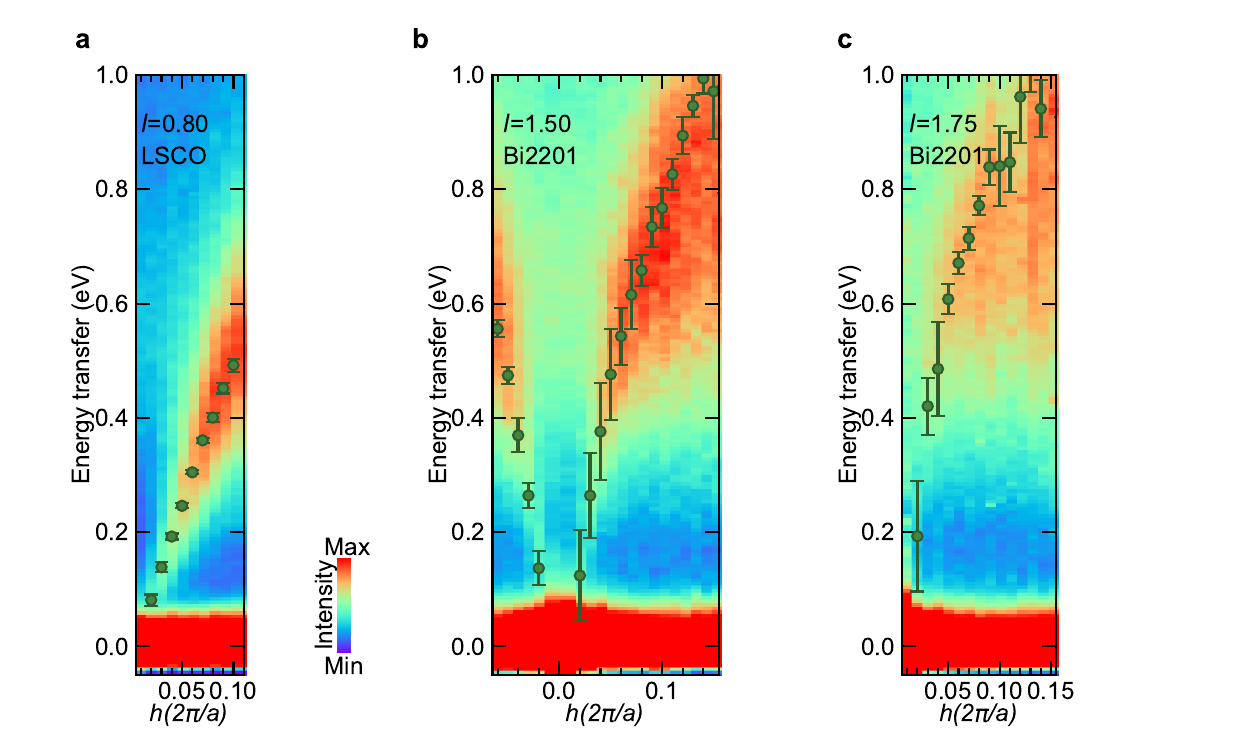}
		\caption{\textbf{Additional in-plane plasmon dispersions in LSCO and Bi2201.}
			\textbf{a} RIXS intensity map of LSCO for momentum transfer along the $h$ direction at $l=0.80$ and \textbf{b, c} of Bi2201 at $l=1.50$ and $l=1.75$. Green circle symbols indicate the least-square-fit peak positions of the plasmon excitations. Fitted plasmon energies for LSCO have been used in Fig. 4(a) of the main paper. 
		} \label{Extended_hscans}
	\end{center}
\end{figure*}

\begin{figure}[htbp!]
	\begin{center}
		\includegraphics{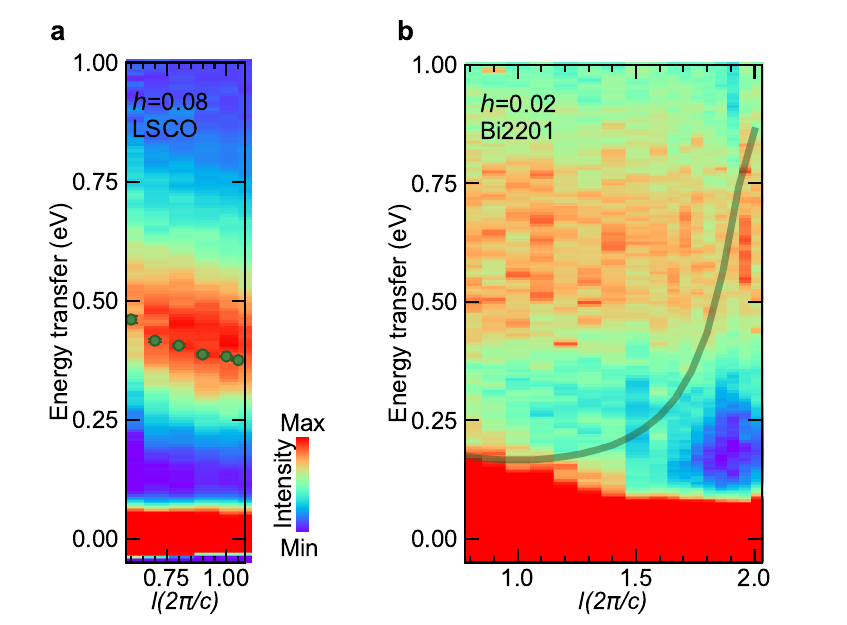}
		\caption{\textbf{Additional out-of-plane plasmon dispersions in LSCO and Bi2201.}
			\textbf{a} RIXS intensity map of LSCO for momentum transfer along the $l$-direction at $h=0.80$ and \textbf{b} of Bi2201 at $h=0.02$. Green circle symbols indicate the least-square-fit peak positions of the plasmon excitations. Error bars  are smaller than the symbols. Continuous line shows the expected plasmon dispersion from $t$-$J$-$V$ model.
		} \label{Extended_lscans}
	\end{center}
\end{figure}

\begin{figure}[hb!]
	\begin{center}
		\includegraphics{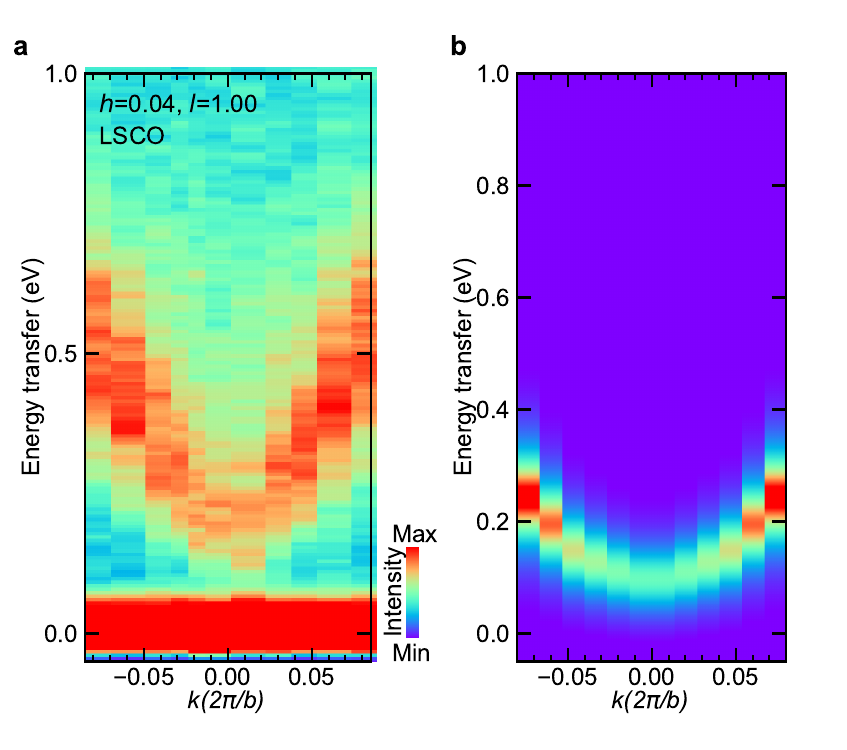}
		\caption{\textbf{Plasmon dispersion in LSCO along $k$-direction.}
			\textbf{a} RIXS intensity map of LSCO for momentum transfer along the $k$ direction at ($h=0.04$, $l=1.0$). \textbf{b}  Intensity map of the charge susceptibility corresponding to the spectra shown in \textbf{a}, calculated using $t$-$J$-$V$ model.
		} \label{Kscan_LSCO}
	\end{center}
\end{figure}

\begin{figure*}[htbp!]
	\begin{center}
		\includegraphics{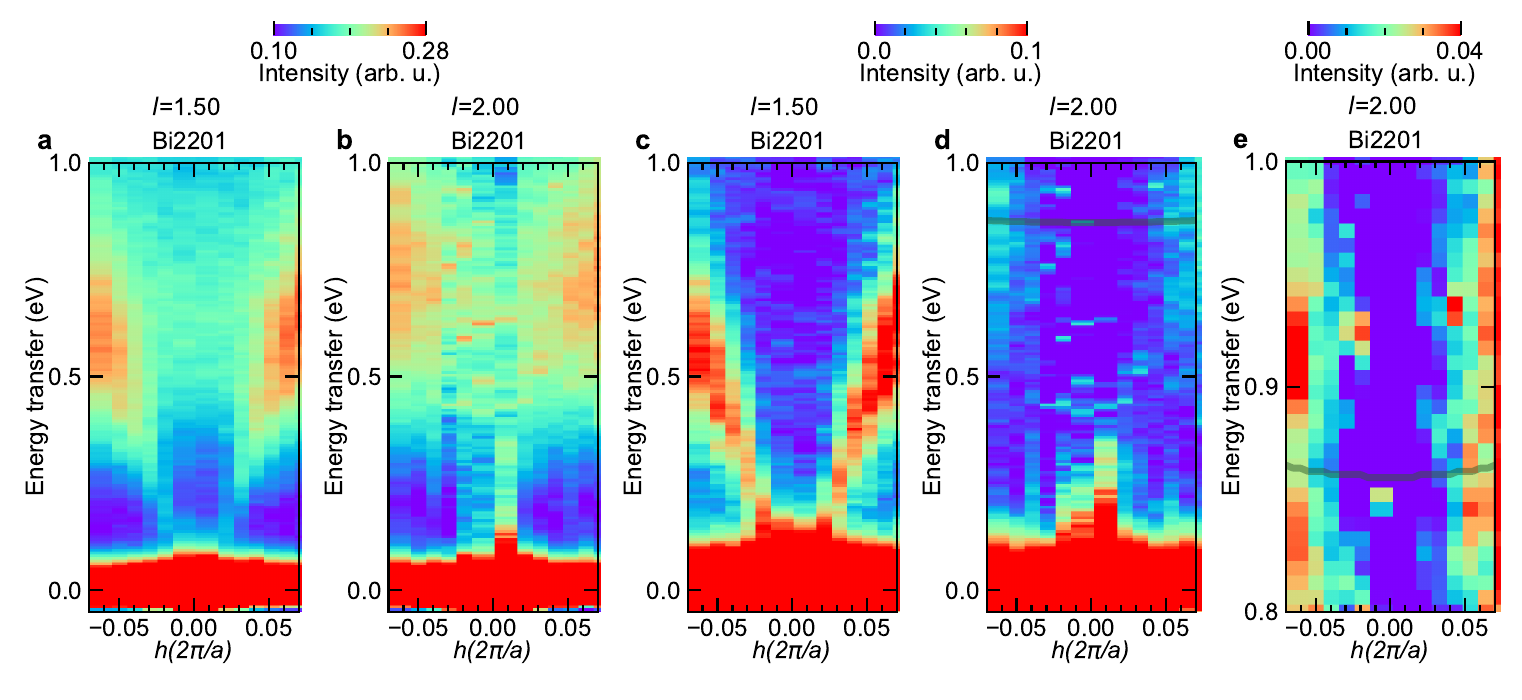}
		\caption{\textbf{Optical plasmon dispersions in Bi2201.}
			\textbf{a, b} RIXS intensity maps of Bi2201 for momentum transfer along the $h$-direction at $l=1.50$ and $l=2.00$. \textbf{c, d} RIXS intensity maps of Bi2201 after subtraction of bimagnon excitations from \textbf{a} and \textbf{b}. \textbf{e} Zoomed in map of \textbf{d}. Continuous line in \textbf{d, e} are the expected optical plasmon dispersion.  For both $h$ close to 0 and $l$ close to 2 the plasmon amplitudes reduce substantially (see Fig.~\ref{Amplitude_and_fwhm}). 
		} \label{Optical_plasmon}
	\end{center}
\end{figure*}

RIXS intensity map of LSCO for momentum transfer along the $h$-direction at $l=0.80$ is shown in Fig.~\ref{Extended_hscans}(a). The plasmon energies from these scans have been used in Fig.~4(a) of the main paper. Extended $h$-scans are shown in Fig.~\ref{Extended_hscans}(b, c) till $h=0.15$, showing the continuous rise of the plasmon energies towards the $dd$ excitations in Bi2201. RIXS intensity map of LSCO for momentum transfer along the $l$-direction at $h=0.08$ is shown in Fig.~\ref{Extended_lscans}(a). The plasmon energies from these scans have been used in Fig.~4(c) of the main paper. In Fig.~\ref{Extended_lscans}(b) RIXS intensity map of Bi2201 for momentum transfer along the $l$-direction at $h=0.02$ is shown. Although plasmon dispersion is visible for $h=0.02$ in Bi2201, the energies could not be extracted by fitting the data, due to either proximity to elastic line below $l=1.5$ or weak spectral weight above $l=1.5$. The over-plotted continuous dispersion line obtained from $t$-$J$-$V$ model seemingly follows the plasmon spectral weight. Plasmon excitations were however observed along the $k$ directions in LSCO (see Fig.~\ref{Kscan_LSCO}(a)) as expected from the 4-fold symmetry of orbitals in the CuO$_2$ planes. As such, using the same parameters of the $t$-$J$-$V$ model optimised for plasmon dispersions along $h$- and $l$-directions, the plasmon excitations along the $k$ direction can be reproduced (see Fig.~\ref{Kscan_LSCO}(b)). Due to the large $c$-axis lattice parameter of Bi2201, we were able to probe close to $l=2$, however, strong suppression of the charge excitations in this region, forbade us from observing the optical branch of plasmons (see Fig.~\ref{Optical_plasmon}). Although some residual spectral weight is visible for slightly higher $h$-values at $l=2.00$, the optical plasmon branch still remains intangible. 

Close to the zone-centre, high energy-resolution scans are need to differentiate the plasmon peak from the elastic and the phonon peaks. This is important if one wants to study the validity of the $t$-$J$-$V$ model with interlayer hopping which predicts a zone-centre gap for the acoustic plasmons. From the current results we can only estimate the upper limit of the acoustic plasmon energies at the zone-centre to be  $\approx0.075$~eV for LSCO, setting interlayer hopping $t_z\lesssim0.007$~eV in this material (see Fig.~\ref{gamma}).


\section{RIXS data fitting}
\noindent
RIXS data were normalised to the incident photon flux, and subsequently corrected for self-absorption effects using the procedure described in~\cite{minola2015prl} prior to fitting. A Gaussian lineshape with the experimental energy resolution was used to fit the elastic line. Gaussian lineshapes were also used to fit the low energy phonon excitations at $\sim 0.045$~eV and their overtones. The scattering intensities $S$($\mathbf{Q}$, $\omega$) of the plasmons and bimagnons at given values of $\mathbf{Q}=ha^*+kb^*+lc^*$ ($a^*=2\pi/a$, $b^*=2\pi/b$, $c^*=2\pi/c$), dependent on the imaginary part of their respective dynamic susceptibilities $\mathbf{\chi}''$($\mathbf{Q}$, $\omega$) were modelled as:
\begin{equation}
	S(\mathbf{Q}, \omega) \propto \frac{\mathbf{\chi}''(\mathbf{Q}, \omega)}{1- e^{-\hbar\omega/k_BT} },
\end{equation}
where $k_B$, $T$ and $\hbar$ are the Boltzmann constant, temperature and the reduced Planck constant. A generic damped harmonic oscillator model can be used for the response function 

\begin{equation}
	\mathbf{\chi}''(\mathbf{Q}, \omega) \propto \frac{\gamma\omega}{\left[\omega^2-\omega^2_0\right]^2+4\omega^2\gamma^2},
	\label{dho}
\end{equation}

\noindent where $\omega_0$ and $\gamma$ are the undamped frequency and the damping factor respectively. Eq.~\ref{dho} can be equivalently written using an anti-symmetrised Lorentzian function,
\begin{equation}
	\frac{1}{\omega_p}\left[\frac{\gamma}{(\omega-\omega_p)^2+\gamma^2}-\frac{\gamma}{(\omega+\omega_p)^2+\gamma^2}\right],
\end{equation}

\noindent with peaks at $\pm\omega_p$ for $\omega_p^2=\omega_0^2-\gamma^2$, given that $\gamma\le\omega_0$, which was found to hold for the plasmon excitations observed in this study (see Fig.~\ref{Amplitude_and_fwhm}). In the results, we plotted the plasmon propagation energy as the peak $\omega_p$ of this function.

First we extracted the zone-centre energy, amplitude and width of the broad incoherent mode at $h=0.01$ and concluding this to be a bimagnon, fixed its amplitude and width for the whole ($h$, $l$)-range~\cite{bisogni2012prb,vernay2007prb}. The energy values of the bimagnons were allowed to vary within $\pm20$ meV for the RIXS spectra along $h$-direction. For the RIXS spectra along $l$-direction, bimagnon energies were kept fixed to the values obtained for corresponding $h$-values from the $h$-direction scans.  This allowed us to decompose the inelastic spectra into two components with less ambiguity, especially for the $h$-values where energies of the two modes were nearby. Significant correlations were however found below $h<0.02$, between the elastic, phonon and plasmon amplitudes and energies, and hence the plasmon energy values determined in these regions are less conclusive.  A high energy quadratic background was also included in the fitting model to account for the tailing contribution from $dd$-excitations above 1.5 eV. 

\begin{figure}[htbp!]
	\begin{center}
		\includegraphics{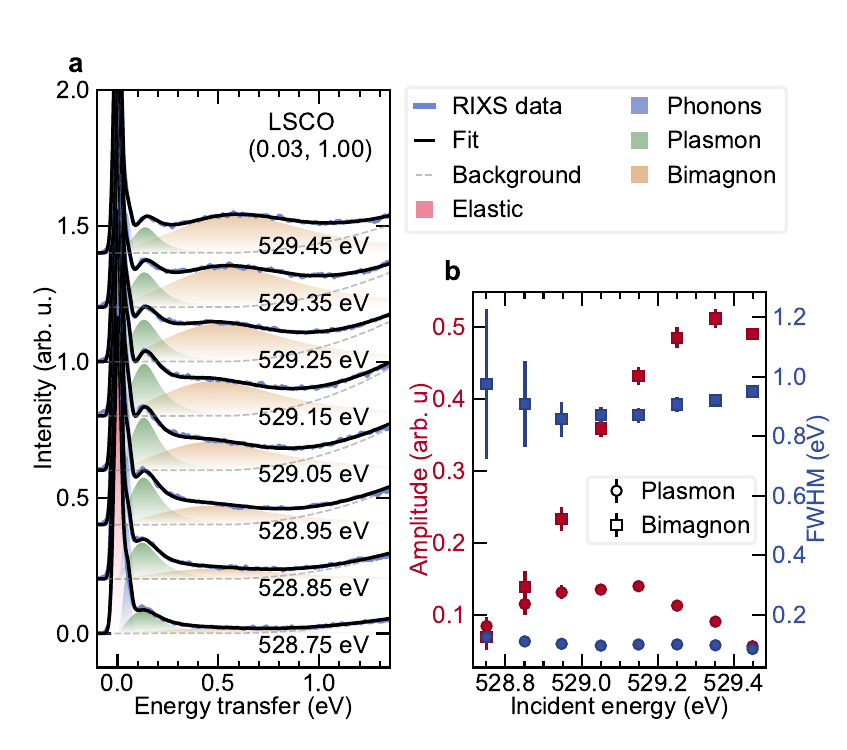}
		\caption{\textbf{RIXS spectra fits of LSCO for varying incident energies.}
			\textbf{a,} Fits to the RIXS spectra of LSCO at ($h=0.03$, $l=1.00$) for mentioned incident energies about the \textit{hole}-peak in O $K$-edge XAS presented in Fig. 1(d) of main paper are shown by using the model described in Section III. The extracted plasmon and bimagnon peak energies have been used in Fig. 1(g) of main paper. \textbf{b,} The plasmon and bimagnon amplitudes (circle and square red symbols) and widths (circle and square blue symbols) as a variation of incident energy is plotted. 
		} \label{LSCO_edfits}
	\end{center}
\end{figure}

\begin{figure*}[htbp!]
	\begin{center}
		\includegraphics{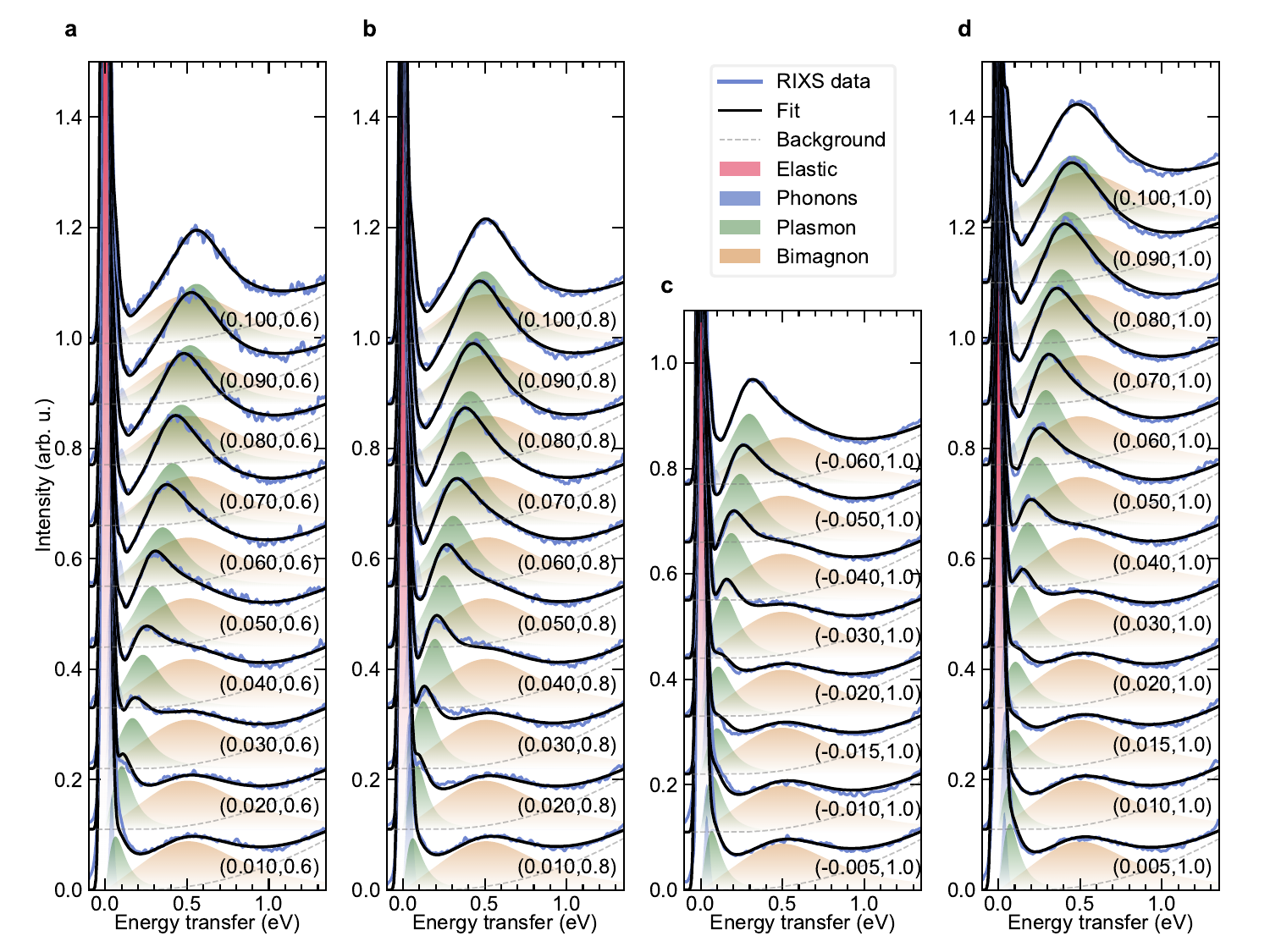}
		\caption{\textbf{RIXS spectra fits of LSCO for momentum transfer along the $h$ direction.}
			\textbf{a, b, c,} and \textbf{d,} Fits to the RIXS spectra of LSCO presented in Fig. 2 (a, b, e) of main paper at mentioned ($h$, $l$)-values are shown by using the model described in Section III. The extracted plasmon peak energies have been used in Fig. 2(a, b, e) and Fig. 4(a) in the main paper and Fig.~\ref{Extended_hscans}. The plasmon amplitudes and widths have been used in Fig.~\ref{Amplitude_and_fwhm}.
		} \label{LSCO_hfits}
	\end{center}
\end{figure*}

\begin{figure*}[htbp!]
	\begin{center}
		\includegraphics{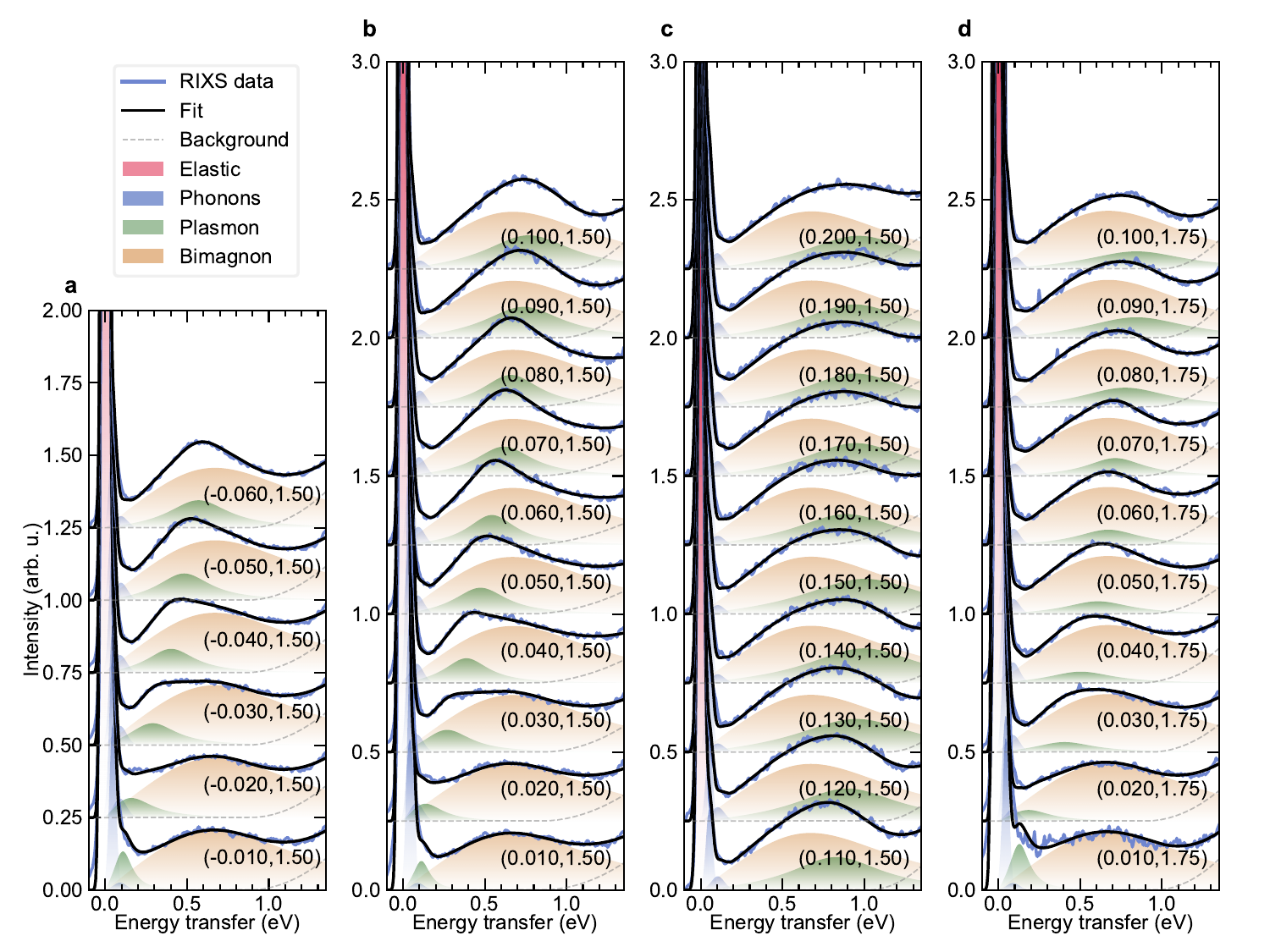}
		\caption{\textbf{RIXS spectra fits of Bi2201 for momentum transfer along the $h$ direction.}
			\textbf{a, b, c,} and \textbf{d,} Fits to the RIXS spectra of Bi2201 presented in Fig. 2 (c, d, f) of main paper at mentioned ($h$, $l$)-values are shown by using the model described in Section III. The extracted plasmon peak energies have been used in Fig. 2(c, d, f) and Fig. 4(a) in the main paper and Fig.~\ref{Extended_hscans}. The plasmon amplitudes and widths have been used in Fig.~\ref{Amplitude_and_fwhm}.
		} \label{Bi2201_hfits}
	\end{center}
\end{figure*}

\begin{figure*}[htbp!]
	\begin{center}
		\includegraphics{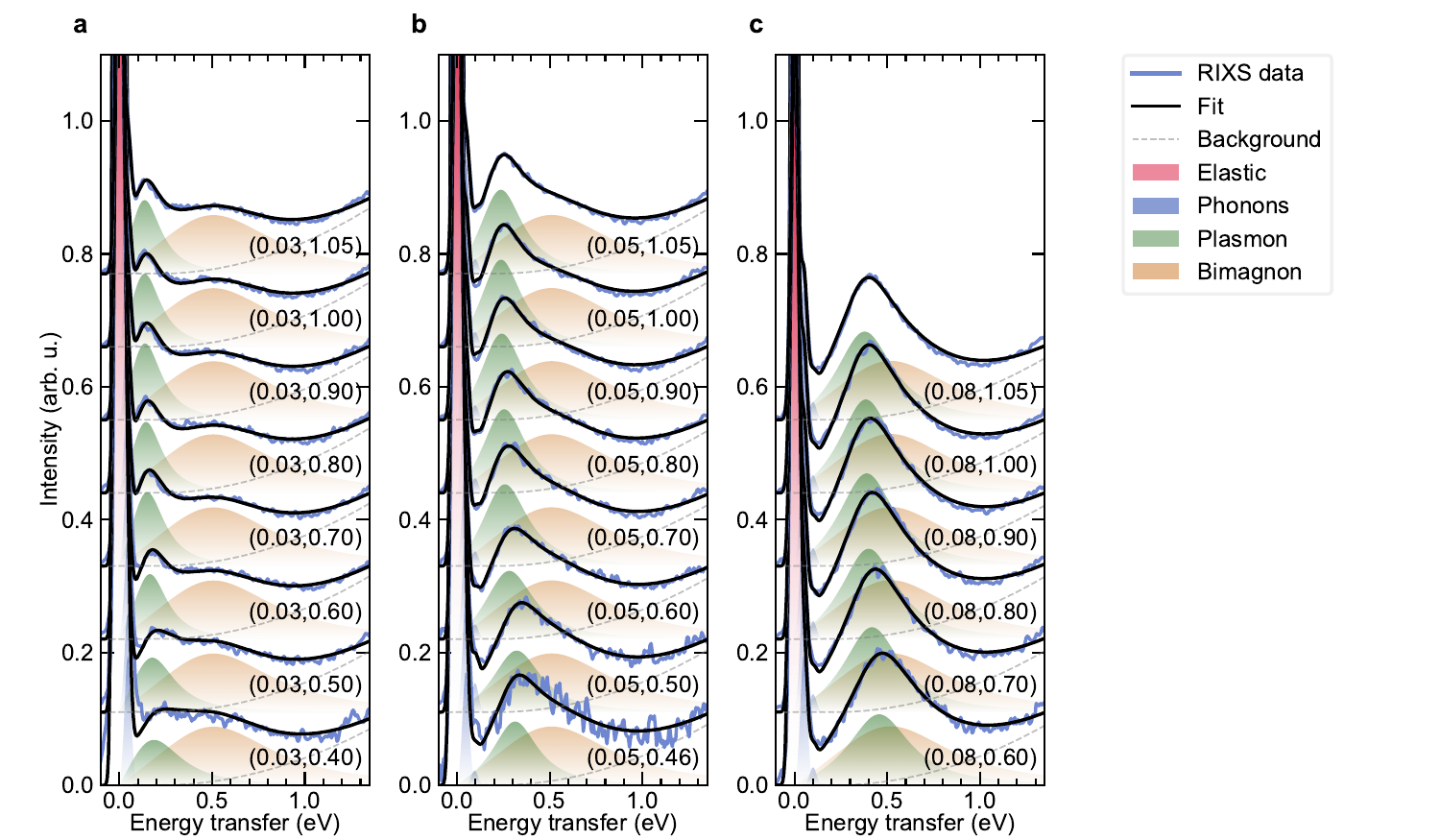}
		\caption{\textbf{RIXS spectra fits of LSCO for momentum transfer along the $l$-direction.}
			\textbf{a, b,} and \textbf{c,} Fits to the RIXS spectra of LSCO presented in Fig. 3 (a, c) of main paper at mentioned ($h$, $l$)-values are shown by using the model described in Section III. The extracted plasmon peak energies have been used in Fig. 3(a, c) and Fig. 4(c) in the main paper and Fig.~\ref{Extended_lscans}. In Fig. 3(e) of the main paper, RIXS spectra subtracted by the fitted bimagnon contribution have been shown. The plasmon amplitudes and widths have been used in Fig.~\ref{Amplitude_and_fwhm}. 
		} \label{LSCO_lfits}
	\end{center}
\end{figure*}

\begin{figure*}[htbp!]
	\begin{center}
		\includegraphics{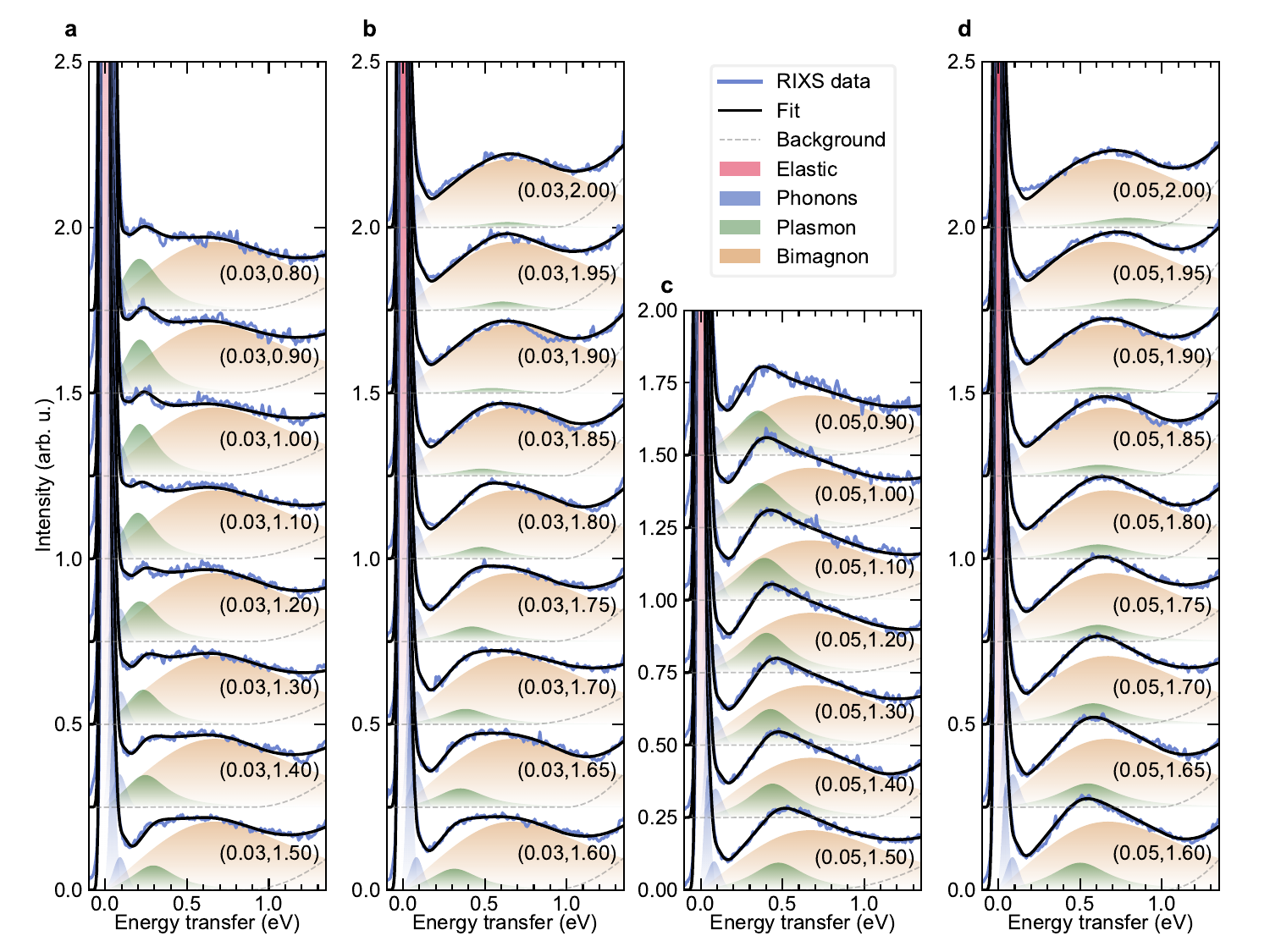}
		\caption{\textbf{RIXS spectra fits of Bi2201 for momentum transfer along the $l$-direction.}
			\textbf{a, b, c,} and \textbf{d,} Fits to the RIXS spectra of Bi2201 presented in Fig. 3 (b, d) of main paper at mentioned ($h$, $l$)-values are shown by using the model described in Section III. The extracted plasmon peak energies have been used in Fig. 3(b, d) and Fig. 4(c) in the main paper. In Fig. 3(e) of the main paper, RIXS spectra subtracted by the fitted bimagnon contribution in this manner have been shown. The plasmon amplitudes and widths have been used in Fig.~\ref{Amplitude_and_fwhm}. 
		} \label{Bi2201_lfits}
	\end{center}
\end{figure*}

\begin{figure*}[htbp!]
	\begin{center}
		\includegraphics{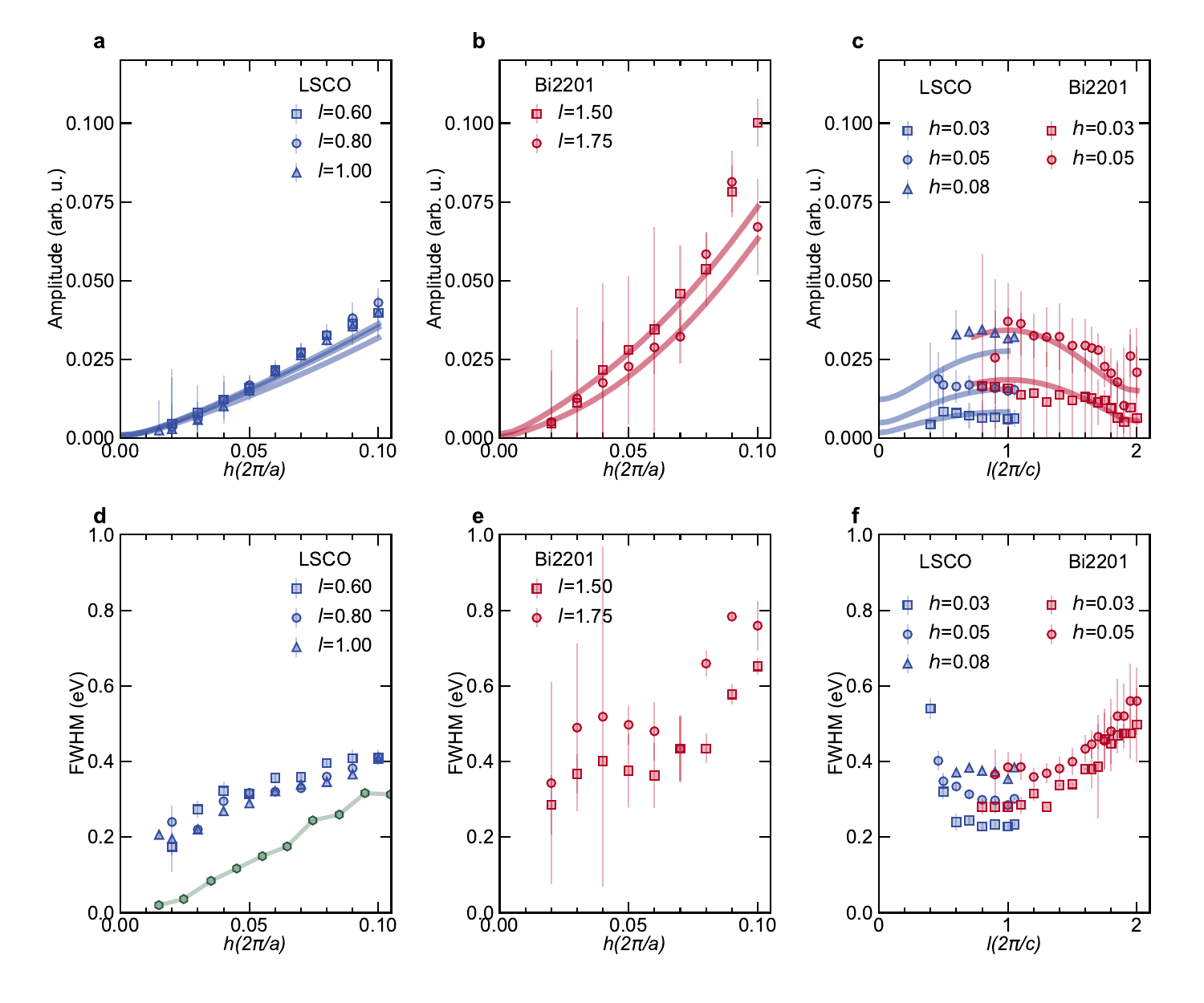}
		\caption{\textbf{a, b,} and \textbf{d, e,}  Plasmon amplitudes and widths in LSCO (blue symbols) and Bi2201 (red symbols) for momentum transfer along the $h$-direction for different $l$-values and \textbf{c,} and \textbf{f,} for momentum transfer along the $l$-direction for different $h$-values, summarised from least-square-fits of RIXS spectra. Continuous lines are integrated spectral weights of $\mathbf{\chi}_c''$($\mathbf{Q}$, $\omega$) calculated within the $t$-$J$-$V$ model. To make an appropriate comparison, all the calculated amplitude values were scaled to the experimental values at $h=0.05, l=1.00$ for LSCO and $h=0.05, l=1.50$ for Bi2201.  Also compared are the plasmon widths from electron-doped La$_{2-x}$Ce$_x$CuO$_4$ ($x$=0.175) (green line with symbols)~\cite{hepting2018nat}.
		} \label{Amplitude_and_fwhm}
	\end{center}
\end{figure*}

The RIXS spectra fits of LSCO for incident energy detuned scans are shown in Fig.~\ref{LSCO_edfits}. Also in Fig.~\ref{LSCO_edfits}(b) are shown the change in plasmon and bimagnon amplitudes and widths as the incident energy is varied. The non-resonant behaviour of the bimagnon amplitude implies an incoherent character in sharp contrast to the plasmon. RIXS spectra fits of LSCO and Bi2201 for momentum transfer along the $h$-direction are shown in Fig.~\ref{LSCO_hfits} and \ref{Bi2201_hfits}, and along the $l$-direction in Fig.~\ref{LSCO_lfits} and \ref{Bi2201_lfits} respectively. Plasmon amplitude and width variation along $h$- and $l$-directions in LSCO and Bi2201 from these fits are summarised in Fig.~\ref{Amplitude_and_fwhm}.

\section{$t$-$J$-$V$ model}
For discussing the nature and origin of three-dimensional (3D) charge excitations in LSCO and Bi2201 we employed the minimal layered $t$-$J$-$V$ model~\cite{greco2019cp,zhang1988prb}:

\begin{equation}
	H = -\sum_{i, j,\sigma} t_{i j}\tilde{c}^\dag_{i\sigma}\tilde{c}_{j\sigma}
	+ \sum_{\langle i,j \rangle} J_{ij} \left( \vec{S}_i \cdot \vec{S}_j - \frac{1}{4} n_i n_j \right)
	+\frac{1}{2} \sum_{i,j}  V_{ij} n_i n_j. 
	\label{tJV} 
\end{equation}

Here, $t_{i j}$ represents the hopping parameter and $J_{i j}$ the exchange parameter.  The 3D form of long-range Coulomb interaction $V_{i j}$ used in Eq.~\ref{tJV} in momentum space is~\cite{becca1996prb}:
\begin{equation}
	V(\mathbf{Q})=\frac{V_c}{A(q_x,q_y) - \cos q_z},
	\label{LRC}
\end{equation}

\noindent where $V_c= e^2 d(2 \epsilon_{\perp} a^2)^{-1}$ and $A(q_x,q_y)= \alpha (2 - \cos q_x - \cos q_y)+1$ 
with $\alpha=\frac{\epsilon_\parallel/\epsilon_\perp}{(a/d)^2}$, $e$ the elementary charge and high frequency in-($\epsilon_{\parallel}$) and out-of-plane ($\epsilon_{\perp}$) dielectric constants.

On each plane the hopping parameter $t_{i j}$ takes a value $t$ $(t')$ between the first (second) nearest-neighbours sites on the square lattice and $J_{i j}$ is the exchange interaction between the nearest-neighbours. Since hole-doped cuprates are correlated electron systems the $t$-$J$ model is believed to be a minimal model of the CuO$_2$ planes~\cite{zhang1988prb}. The fact that we deal with a correlated system is contained in $\tilde{c}^\dag_{i\sigma}$ and $\tilde{c}_{i\sigma}$ which are the creation and annihilation operators, respectively, of electrons with spin $\sigma(=\uparrow, \downarrow)$ in the Fock space without any double occupancy. $n_i$ is the electron density operator and $\vec{S}_i$ the spin operator. The 3D nature of the model originates from the presence of a hopping $t_z$ between the adjacent planes, and the long-range Coulomb interaction $V_{ij}$ for a layered system. The form of $V_{i j}$ in Eq.~\ref{tJV} in momentum space is given in Eq.~\ref{LRC}~\cite{becca1996prb}.  Finally, the indices $i$ and $j$ run over the sites of a three-dimensional lattice, and $\langle i,j \rangle$ indicates a pair of nearest-neighbour sites.

A theoretical treatment of this model is non-trivial because the Hamiltonian is defined in the restricted  Hilbert space that prohibits non-double occupancy on each site, which complicate the commutation rules of the operators. In addition, there is no small parameter for perturbation. We implement a large-$N$ expansion~\cite{foussats2004prb,greco2016prb,greco2019cp} where the spin index $\sigma$ is extended to a new index $p$ running from $1$ to $N$. In order to get a finite theory in the limit $N\rightarrow \infty$, we rescale the hopping $t_{ij}$ to $t_{ij}/N$, $J$ to $J/N$ and $V_{ij}$ to $V_{ij}/N$, and $1/N$ is used as the small parameter to control the expansion. $N$ is put to $N=2$ in the end. Although the physical value is $N=2$, the large-$N$ expansion has several advantages over usual perturbations theories. Applying the large-$N$ treatment~\cite{greco2016prb} the quasiparticles disperse in momentum space as 

\begin{equation}
	\epsilon_{\mathbf{k}} = \epsilon_{\mathbf{k}}^{\parallel}  + \epsilon_{\mathbf{k}}^{\perp} 
	\label{Ek}
\end{equation}

\noindent where the in-plane $\epsilon_{\mathbf{k}}^{\parallel}$ and the out-of-plane $\epsilon_{\mathbf{k}}^{\perp}$ dispersions are given by, respectively,  

\begin{equation}
	\epsilon_{\mathbf{k}}^{\parallel} = -2 \left( t \frac{\delta}{2}+\Delta \right) (\cos k_{x}+\cos k_{y})
	- 4t' \frac{\delta}{2} \cos k_{x} \cos k_{y} - \mu 
	\label{Epara}
\end{equation}

and
\begin{equation}
	\epsilon_{\mathbf{k}}^{\perp} = 2 t_{z} \frac{\delta}{2} (\cos k_x-\cos k_y)^2 \cos k_{z}. 
	\label{Eperp}
\end{equation}

The functional form $ (\cos k_x-\cos k_y)^2$ in $\epsilon_{\mathbf{k}}^{\perp}$ is frequently invoked for cuprates~\cite{andersen1995jpcs}. Other forms for $\epsilon_{\mathbf{k}}^{\perp}$, however, do not change the qualitative features. Although the electronic dispersion looks similar to that in a free electron system, the hopping integrals $t$, $t'$, and $t_z$ are renormalised by doping $\delta$ because of electron correlation effects. For both the optimally-doped materials we use $\delta=0.16$.

The term $\Delta$ in Eq.~\ref{Epara}, which is proportional to $J$, is the mean-field value of the bond variables introduced to decouple the exchange term through a  Hubbard-Stratonovich transformation~\cite{foussats2004prb,greco2016prb}. The value of $\Delta$ is computed self-consistently together with the chemical potential $\mu$ for a given $\delta$ by using 

\begin{equation}
	\Delta = \frac{J}{4{N_sN_z}} \sum_{\mathbf{k}, \eta} \cos(k_\eta) n_F(\epsilon_{\mathbf{k}}), 
	\label {Delta}
\end{equation}
and 
\begin{equation}
	(1-\delta)=\frac{2}{N_s N_z} \sum_{\mathbf{k}} n_F(\epsilon_{\mathbf{k}}),
\end{equation}

\noindent where $n_F$ is the Fermi function, and $N_s$ and $N_z$ are the total number of lattice sites on the square lattice and the number of layers along the $c$ direction respectively. We take the number of layers $N_z$ equal to 30, which should be large enough, and 
set the temperature to zero.

As shown previously~\cite{foussats2004prb,greco2016prb,greco2019cp}, the charge-charge correlation function  
$\mathbf{\chi}_c (\mathbf{r}_i -\mathbf{r}_j, \tau)=\left\langle T_\tau n_i(\tau) n_j(0)\right\rangle $ can be 
computed in the $\mathbf{q}$-$\omega$ space as 

\begin{eqnarray}
	\mathbf{\chi}_c(\mathbf{q},\omega)= N \left ( \frac{\delta}{2} \right )^{2} D_{11}(\mathbf{q},\omega).
\end{eqnarray}

\noindent Thus, $\mathbf{\chi}_c$ is the element $(1,1)$ of the $6 \times 6$ bosonic propagator $D_{ab}$ where

\begin{equation}
	D^{-1}_{ab}(\mathbf{q},\mathrm{i}\omega_n)
	= [D^{(0)}_{ab}(\mathbf{q},\mathrm{i}\omega_n)]^{-1} - \Pi_{ab}(\mathbf{q},\mathrm{i}\omega_n),
	\label{dyson}
\end{equation}

\noindent and the matrix indices $a$ and $b$ run from 1 to 6. 
$D^{(0)}_{ab}(\mathbf{q},\mathrm{i}\omega_n)$ is a bare bosonic propagator

\begin{equation} \label{D0inverse}
	[D^{(0)}_{ab}(\mathbf{q},\mathrm{i}\omega_n)]^{-1} = N 
	\left(
	\begin{array}{llllll}
		\frac{\delta^2}{2} \left[ V(\mathbf{q})-J(\mathbf{q})\right] 
		& \frac{\delta}{2} & 0 & 0 & 0 & 0\\
		\frac{\delta}{2} & 0 & 0 & 0 & 0 & 0\\
		0 & 0 & \frac{4\Delta^2}{J} & 0 & 0 & 0\\
		0 & 0 & 0 & \frac{4\Delta^2}{J} & 0 & 0\\
		0 & 0 & 0 & 0 & \frac{4\Delta^2}{J} & 0\\
		0 & 0 & 0 & 0 & 0 & \frac{4\Delta^2}{J}
	\end{array}
	\right),
\end{equation}

\noindent and $\Pi_{ab}$ are the bosonic self-energies,
\begin{eqnarray}
	&& \Pi_{ab}(\mathbf{q},\mathrm{i}\omega_n)
	= -\frac{N}{N_s N_z}\sum_{\mathbf{k}} h_a(\mathbf{k},\mathbf{q},\varepsilon_\mathbf{k}-\varepsilon_{\mathbf{k}-\mathbf{q}}) 
	\frac{n_F(\varepsilon_{\mathbf{k}-\mathbf{q}})-n_F(\varepsilon_\mathbf{k})}
	{\mathrm{i}\omega_n-\varepsilon_\mathbf{k}+\varepsilon_{\mathbf{k}-\mathbf{q}}} 
	h_b(\mathbf{k},\mathbf{q},\varepsilon_\mathbf{k}-\varepsilon_{\mathbf{k}-\mathbf{q}}) \nonumber \\
	&& \hspace{25mm} - \delta_{a\,1} \delta_{b\,1} \frac{N}{N_s N_z}
	\sum_\mathbf{k} \frac{\varepsilon_\mathbf{k}-\varepsilon_{\mathbf{k}-\mathbf{q}}}{2}n_F(\varepsilon_\mathbf{k}) \; , 
	\label{Pi}
\end{eqnarray}

\noindent where the six-component vertex $h_a$ is given by

\begin{equation}
	\begin{split}
		h_a(\mathbf{k},\mathbf{q},\nu) = \left\{
		\frac{2\varepsilon_{\mathbf{k}-\mathbf{q}}+\nu+2\mu}{2}+
		2\Delta \left[ \cos\left(k_x-\frac{q_x}{2}\right)\cos\left(\frac{q_x}{2}\right) +
		\cos\left(k_y-\frac{q_y}{2}\right)\cos\left(\frac{q_y}{2}\right) \right];1;
		\right. \nonumber \\
		\left. -2\Delta \cos\left(k_x-\frac{q_x}{2}\right); -2\Delta \cos\left(k_y-\frac{q_y}{2}\right);
		2\Delta \sin\left(k_x-\frac{q_x}{2}\right);  2\Delta \sin\left(k_y-\frac{q_y}{2}\right)
		\right\}.
	\end{split}
\end{equation}

\noindent Here $\mathbf{q}$ and $\mathbf{k}$ are  three dimensional wavevectors and $\omega_n$ is a bosonic Matsubara frequency.
The factor $N$ comes from the sum over the $N$ fermionic channels. The $6$ channels involved in $D_{ab}(\mathbf{q},\mathrm{i}\omega_n)$ come from on-site charge fluctuations, fluctuations of a Lagrangian multiplier introduced to impose the non-double occupancy at any site, and fluctuations of the four bond variables which depend on $J$~\cite{greco2016prb,foussats2004prb}. 

To describe plasmon excitations we compute the spectral weight of the density-density correlation function Im$\mathbf{\chi}_c(\mathbf{q},\omega)$ after analytical continuation 
\begin{equation}
	{\rm i} \omega_n \rightarrow \omega + {\rm i} \Gamma.
	\label{continuation}
\end{equation}

\begin{figure*}[htbp!]
	\begin{center}
		\includegraphics{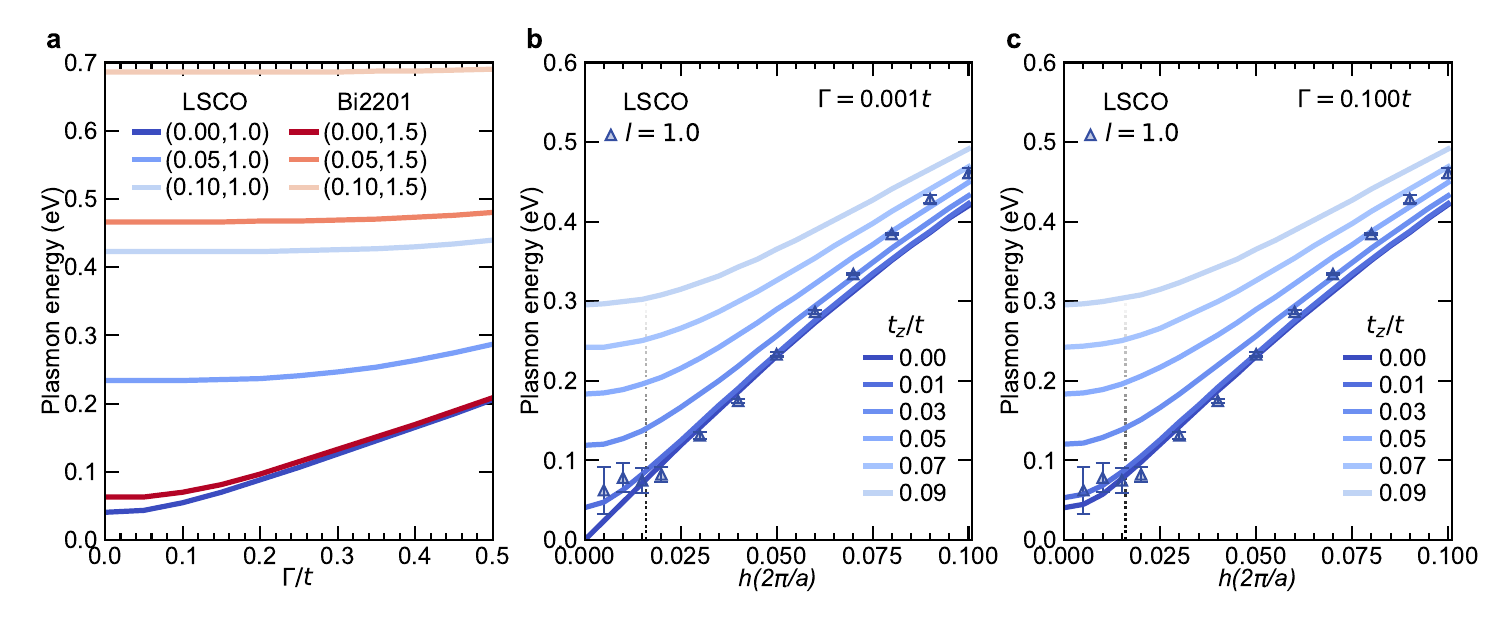}
		\caption{\textbf{Effect of interlayer hopping on gap formation in acoustic plasmon branches}
			\textbf{a,} Continuous lines are plasmon energies calculated for optimised parameters for LSCO and Bi2201 at given ($h$, $l$)-values as a function of broadening parameter $\Gamma$. \textbf{b,} and \textbf{c,} The effect of interlayer hopping parameter $t_z$, on the plasmon energy gap at $h=0.00$, $l=1.0$ for $\Gamma=0.001t$ and $\Gamma=0.1t$ respectively. For comparison, fitted plasmon energy values from RIXS spectra collected on LSCO for $l=1.00$ is also shown. Two values of $\Gamma$ are shown since it affects the plasmon energies close to the zone-center for the acoustic branches as shown in panel (a).  Black dotted lines are $h$-values below which plasmon energies are not reliable due to large correlation with the elastic and phonon intensities.
		} \label{gamma}
	\end{center}
\end{figure*}

\noindent Since $\Gamma$ influences the width of the plasmon, its effect on the plasmon peak position is strongest when it becomes comparable to undamped plasmon energy (overdamped condition). As observed in the experiment (see Fig.~\ref{Amplitude_and_fwhm}(d, e) and (f)), this condition may only be true close to the zone-centre for the acoustic plasmons. Plasmon energies calculated from the optimised models, as function of $\Gamma$, plotted in Fig.~\ref{gamma}(a) clearly demonstrates that plasmon dispersions have negligible effect at larger $h$ values. Here a small positive value $\Gamma=0.1t$ was chosen for all the plasmon dispersion simulations. However to replicate the plasmon peak profiles observed in Fig.~3(f) of main paper, $\Gamma$ values of $0.2t$ and $0.29t$ were chosen for LSCO and Bi2201 respectively. A finite value of $\Gamma$ contains information of an extrinsic broadening due to the instrumental resolution, and an intrinsic broadening due to incoherent effects due to electronic correlations~\cite{prelovsek1999prb}.

The model Hamiltonian Eq.~\ref{tJV} contains several material dependant parameters. To reduce the number of tuning parameters in the model, we have used, if available, the most common values from literature. The exchange interaction $J$ is considered only inside the plane. The out-of-plane exchange term is much smaller than $J$~\cite{theo1988prb}. The $J$ was taken as $0.3t$ for both the materials given their similar values in the parent compounds~\cite{hybertsen1990prb,peng2017np,lee2006rmp}. It is important to mention here that plasmons are nearly unaffected by the value of $J$. The optical plasmon ($\omega_{\textrm{opt}}$) energy was fixed at 0.85~eV for both LSCO and Bi2201 as reported from optical measurements which are sensitive to only the $l=0$ momentum transfers~\cite{heumen2009njp,uchida1991prb,suzuki1989prb}.  The nearest neighbour hopping parameter $t/2$ was fixed at 0.35~eV for both the materials~\cite{ishii2017prb,horio2018prl}. We took the second-nearest neighbour hopping parameter $t'$, whose exact value is not known apart from reports that $|t'_{\textrm{LSCO}}|<|t'_{\textrm{Bi2201}}|$, to be $-0.2t$ for LSCO and $-0.35t$ for Bi2201~\cite{hashimoto2008prb,pavarini2001prl,greco2019cp,horio2018prl}. We however found that the overall nature of plasmon dispersions does not depend on a precise choice of the band parameters.

We found the optical plasmon energy to be proportional to $\sqrt{V_c/\alpha}$ of Eq.~\ref{LRC}. Depending on the band parameters, we obtained the proportionality constant to be 0.41 for LSCO and 0.40 for Bi2201. With the band parameters fixed, the value of $V_c/\alpha$ was tuned to get $\omega_{\textrm{opt}}=0.85$~eV. Next, $V_c$ and $\alpha$ values were optimised to best match the plasmon dispersions observed in the experiment. For LSCO (Bi2201), we obtained $V_c$ and $\alpha$ to be 18.9~eV (52.5~eV) and 3.47 (8.14) respectively. This gave for LSCO (Bi2201), $\epsilon_\perp$ as $2.21\epsilon_0$ ($1.43\epsilon_0$) and $\epsilon_\parallel$ as $1.62\epsilon_0$ ($1.14\epsilon_0$). Microscopically therefore the relatively large values of $V_c$ and $\alpha$ in Bi2201 come mainly from a large interlayer spacing $d$. The interlayer hopping $t_z$ was tuned to match the plasmon energies for $h\rightarrow0.0$ and an upper limit of $0.01t$ was found for LSCO (see Fig.~\ref{gamma}).

\section{Comparison of plasmon energies between LSCO and Bi2201}
\begin{figure}[htbp!]
	\begin{center}
		\includegraphics{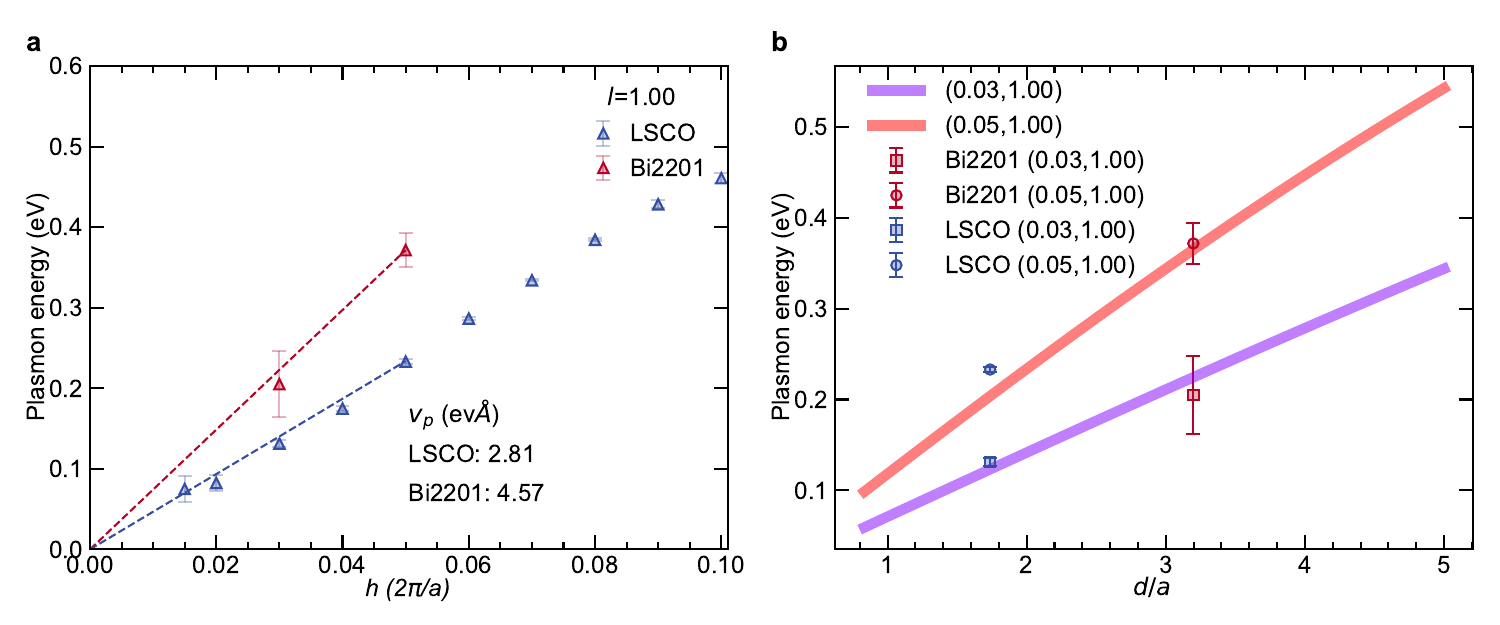}
		\caption{\textbf{Comparison of plasmon energies in LSCO and Bi2201 at $l=1$.}
			\textbf{a}, LSCO and Bi2201 plasmon energies at $l = 1.00$ from experimental RIXS spectra and liner fits to  the extracted $h$-direction plasmon energies. \textbf{b}, The solid lines are plasmon energies calculated using Eq.~\ref{RPA_plasmon}, for shown values of $h$ and $l$ as a function of interlayer spacing. The value of  $\frac{n e^2}{2 m \varepsilon_0 \varepsilon_{\infty} }$ has been taken as 0.57 for a simplistic match to the experimentally obtained results on the two samples.
		} \label{velocities}
	\end{center}
\end{figure}

\noindent
We observe that the plasmon energies of Bi2201 is much larger than that in LSCO close to $l=1$ (see Fig.~3 and Fig.~4(b) of main paper). Acoustic plasmon velocities ($v_p$) for LSCO and Bi2201 at $l=1.00$ estimated using linear fits to extracted $h$-direction plasmon energies are shown in Fig.~\ref{velocities}(a). The Bi2201 to LSCO $v_p$ ratio is found to be around 1.6, roughly matching the ratio of the interlayer spacing between them ($\sim$1.88). The difference in the ratios could be due to the slightly different dielectric constants of the two materials (see Section IV), given the identical level of doping in the two systems. The larger interlayer spacing essentially enhances the 3D Coulomb interaction (Eq.~2 of main paper) of Bi2201 compared to LSCO and thereby raises its plasmon energies. Our $t$-$J$-$V$ model captures the effect of this interlayer spacing as can be seen from the theoretically calculated solid lines in Fig.~4(b) of main paper. This can also be viewed from the analytical form of plasmon energies of layered electron gases for small in-plane momenta~\cite{giuliani2005book,turkalov2003prb},
\begin{equation}
	\omega_p^2 =   \frac{n e^2 d}{2 m \varepsilon_0 \varepsilon_{\infty} } \left[  \frac{ q_{\|} \sinh(q_{\|} d)}{\cosh(q_{\|} d)-\cos(q_\perp d)}  \right]
	\label{RPA_plasmon}
\end{equation} 
\noindent where, $\omega_p$ is the plasmon energy, $n$ is the doping, $d$ is the interlayer spacing, $\varepsilon_{\infty}$ is the high frequency dielectric constant, $q_\parallel$ and $q_\perp$ are the in- and out-of-plane momentum transfer values respectively. As shown in Fig.~\ref{velocities}(b), the plasmon energies are strongly dependent on interlayer spacing, considering other parameters have similar values.



\begin{figure}[htbp!]
	\begin{center}
		\includegraphics{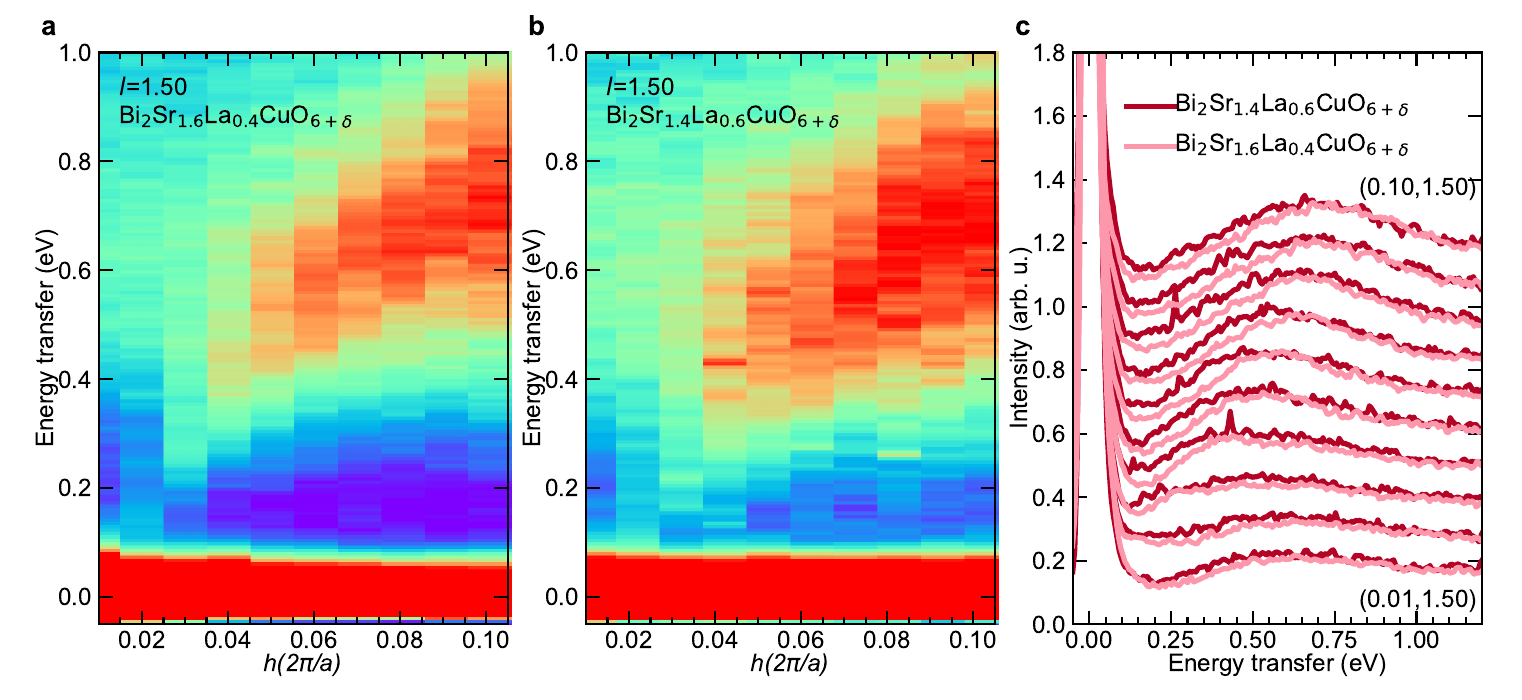}
		\caption{\textbf{Doping dependence of plasmon spectral weight.}
			RIXS intensity maps of \textbf{a},  Bi$_2$Sr$_{1.6}$La$_{0.4}$CuO$_{6+\delta}$ ($p=0.16$)and \textbf{b},  Bi$_2$Sr$_{1.4}$La$_{0.6}$CuO$_{6+\delta}$ ($p=0.14$) for momentum transfer along the $h$-direction at $l = 1.50$. \textbf{c}, Vertically stacked RIXS spectra from $h = 0.01$ to $h = 0.10$ used for the maps in panels \textbf{a} and \textbf{b}.
		} \label{P4_P6}
	\end{center}
\end{figure}

\end{document}